\documentclass[reprint,superscriptaddress,preprintnumbers,nofootinbib,nobibnotes,aps,prd]{revtex4-2}
\pdfoutput=1

\usepackage{graphicx}
\usepackage[dvipsnames]{xcolor}
\usepackage{epsfig}
\usepackage{subfigure}
\usepackage{hyperref}
\usepackage{cancel}
\usepackage{multirow}
\usepackage{parskip}
\usepackage{slashed}
\usepackage{amsmath,amssymb,amstext,amsbsy,amsfonts,amsthm}
\usepackage{svg}
\usepackage{soul}
\usepackage{ulem}

\newcommand{\dd}{\mathrm{d}}

\definecolor{bluetto}{HTML}{0088ff}

\newcommand{\kcl}{Theoretical Particle Physics and Cosmology Group, Physics Department, King's College London, Strand, London WC2R 2LS, United Kingdom}
\newcommand{\ITPH}{Institute for Theoretical Physics, Leibniz University Hannover, Appelstraße 2, 30167 Hannover, Germany}
\newcommand{\MPI}{Max Planck Institute for Gravitational Physics,
Albert Einstein Institute, 30167 Hannover, Germany}

\DeclareMathOperator{\sgn}{sgn}
\DeclareMathOperator{\arsinh}{arsinh}

\hypersetup{
  colorlinks   = true,
  urlcolor     = blue,
  linkcolor    = blue,
  citecolor    = blue
}

\begin{document}

\title{The cosmology of long range Yukawa interactions in general  backgrounds}

\author{Guillem Domènech}
\email{guillem.domenech@itp.uni-hannover.de}
\affiliation{\ITPH}
\affiliation{\MPI}

\author{Panagiotis Giannadakis}
\email{panagiotis.giannadakis@kcl.ac.uk}
\affiliation{\kcl}
\begin{abstract}
Long-range forces in the early universe may lead to early structure formation and, perhaps, to primordial black holes. We generalise previous studies of fermions coupled to a light scalar field by considering general scalar-field-dependent couplings in cosmological backgrounds with a constant equation of state. We identify two broad regimes: a scaling regime, in which the scalar field oscillates around a point of vanishing fermion mass, and an asymptotic regime, in which the field evolves toward configurations where the fermions recover their bare mass. We show that the scaling regime arises from an approximate scale invariance in the scalar-fermion action, which becomes an approximate conformal invariance at late times. In the scaling regime, the ratio between the scalar and fermion energy densities is approximately constant. Our work provides a first step toward a general study of the growth of perturbations in this system.
\end{abstract}

\maketitle

\section{Introduction}

Primordial black holes (PBHs) are promising dark matter candidates and can provide a powerful probe of small-scale physics in the early Universe \cite{Sasaki_2018,Carr_2020,shankaranarayanan2026primordialblackholesreview}. 
In the standard scenario, PBHs are generated from the collapse of local overdensities in the primordial density field \cite{Zeldovich:1967lct,Hawking:1971ei,Carr:1974nx}. However, large density contrasts are required to trigger gravitational collapse in the radiation-dominated universe. Thus, one needs a large variance of the random Gaussian primordial density field. This means that PBH formation is an extremely rare event. 

A common proposed mechanism for generating a sufficient abundance of primordial black holes is to consider inflationary models with a phase of ultra-slow-roll, producing an enhancement of the curvature power spectrum (see, e.g.,  Ref.~\cite{Ozsoy:2023ryl} for a review), although some may argue that severe fine-tuning is required \cite{Azhar:2018lzd,Hertzberg:2017dkh,Nakama:2018utx,Carr:2019hud,Animali:2022otk,Qin:2023lgo,Braglia:2022phb,Cole:2023wyx}. It has been recently argued, that this might be different in a period of kinetic domination, so-called ``kination'' \cite{gouttenoire2022kinationcosmologyscalarfields}, where the universe is dominated by a stiff fluid \cite{cheng2026nonlineardynamicsprimordialblack}. The reason is that, during kination, perturbations decay more slowly than the background stiff fluid. This can generate large local density contrasts, which subsequently undergo nonlinear dynamics, as confirmed by numerical simulations~\cite{cheng2026nonlineardynamicsprimordialblack}.

An alternative mechanism that may generate PBHs is the presence of long-range forces stronger than gravity \cite{Amendola:2017xhl,Flores:2020drq}, often acting only on the dark sector. Although it is still unclear whether PBH formation eventually occurs, long-range forces may substantially boost early structure formation in the radiation-dominated universe \cite{Amendola_2018,Dom_nech_2023}. For instance, the presence of a light scalar field coupled to dark matter can mediate an attractive force, enhancing the growth of dark matter perturbations and potentially leading to halo formation, gravitational collapse, and primordial black hole formation. Very early structure formation may explain, or be constrained, by the early galaxies seen by the James Webb Space Telescope \cite{Haiman_2012,Boylan-Kolchin:2022kae,colazo2024structureformationprimordialblack,hirano2024earlystructureformationprimordial}, the EDGES 21cm signal \cite{Bowman:2018yin} and the seeds of supermassive black hole formation \cite{Inayoshi:2019fun,Lovell_2022,Dayal:2024zwq}.

Long-range scalar forces have been widely considered in dark-sector cosmology \cite{Farrar_2004,Gubser:2004uh,Nusser:2004qu,Bean:2008ac,Archidiacono:2022iuu,Graham:2025gtd}. A characteristic example is quintessence \cite{Wetterich_1988,Wetterich:1994bg,Amendola:1999er,Farrar:2003uw} coupled to a non-relativistic dark matter fluid, for which the scalar dependence of the particle mass induces an effective time-dependent mass for the dark matter particles. Related scalar-dependent mass scenarios also appear in mass-varying-neutrino and growing-neutrino quintessence models
\cite{Fardon:2003eh,Afshordi:2005ym,Wetterich:2007kr,Amendola:2007yx,Casas:2016duf}. One realisation of this idea is a fermion fluid coupled to a scalar through a Yukawa-type interaction. In such models, the scalar mediates an attractive force between fermions, while the fermion mass becomes a function of the scalar background.

The scalar force is long-range whenever the scalar mass is small compared to the physical scale of interest. In this regime, the scalar dynamics is controlled not only by its bare potential, but also by the effective potential induced by the interaction with the fermion fluid. For a scalar-dependent fermion mass, the system can be dynamically driven toward configurations that minimise the effective fermion mass. When this happens, an initially non-relativistic description may break down, as the effective mass can become comparable to, or smaller than, the typical fermion momentum. A fully relativistic formulation is therefore needed.

Such a general relativistic treatment was developed in Ref.~\cite{Dom_nech_2021}, where the dynamics of a homogeneous scalar coupled to a degenerate fermion gas were studied during radiation domination under the assumption of thermal equilibrium. The resulting background solutions provide the starting point for studying perturbations. In Ref.~\cite{Dom_nech_2023}, it was shown that scalar-mediated Yukawa forces can lead to the rapid growth of fermion perturbations in the radiation era, resulting in the formation of dark matter halos; this behaviour was also confirmed using $N$-body numerical simulations. These results suggest that long-range forces in a dark sector may provide a more ``natural'' mechanism for producing large small-scale overdensities--that is, apparently less fine-tuned than the standard scenario, although further investigation is needed to make any definitive claim--and motivate further investigation from the point of view of primordial black hole formation \cite{Flores:2020drq,Flores:2023zpf,Kawana:2021tde,Lu:2024jzs}.

However, the cosmological background expansion may also play an important role in this mechanism. If the post-inflationary Universe is not immediately radiation dominated but instead undergoes an early matter-dominated or kination era, as motivated by non-standard cosmologies and string-theory scenarios~\cite{apers2024stringtheoryhalfuniverse,gouttenoire2022kinationcosmologyscalarfields,Mosny:2025cyd} (see, e.g., Ref.~\cite{Allahverdi:2020bys} for a review), the growth of perturbations and the evolution of the scalar-mediated force can be substantially altered.
Another question concerns the functional form of the scalar-fermion coupling. Most studies have focused either on the linear Yukawa coupling~\cite{Flores:2020drq,Dom_nech_2021,Dom_nech_2023} or the dilatonic coupling~\cite{Jordan1959,BransDicke1961,Fujii:2003pa,Brax:2010gi,Amendola_2018}. However, more general scalar dependencies can be motivated by high-energy extensions. Dilatonic couplings can arise in string theory and modified-gravity scenarios, whereas higher-order monomial couplings can arise as EFT extensions of the Standard Model of particle physics~\cite{baumann2014inflationstringtheory}.

In this work, we restrict ourselves to the homogeneous dynamics of the coupled scalar–fermion system, leaving the analysis of perturbations for future work. We generalise the background analysis of Ref.~\cite{Dom_nech_2021} by considering monomial and dilatonic scalar–fermion couplings on a fixed FLRW background with an arbitrary constant equation-of-state parameter $w$. The scalar-fermion sector is treated as a spectator sector, so that the background expansion is externally prescribed, while the scalar field and the fermion fluid exchange energy through the field dependence of the effective fermion mass.

We present an exhaustive, general analysis of the dynamics of the scalar-fermion system across different cosmological backgrounds and coupling types. Our main result is that the system admits scaling solutions for both classes of couplings and for general cosmological backgrounds. We show that the origin of scaling solutions in the Yukawa interaction case is due to an approximate scale invariance, while in the dilatonic case, the equations of motion admit ``tracking'' solutions, namely when the scalar field derivative is a function of the background expansion rate. Depending on the coupling, the background cosmic fluid, and the presence of a bare scalar potential, the field may oscillate around a saturation point or approach an asymptotic branch. 

This paper is organised as follows. In Sec.~\ref{sec:scalar-fermion}, we introduce the model and provide a conformal-frame argument for the scaling solutions. In Sec.~\ref{sec:Fermi}, we specialise to a degenerate Fermi gas, and in Sec.~\ref{sec:representative_couplings}, we present the general equations of motion for representative couplings. In Secs.~\ref{sec:monomial_results} and~\ref{sec:monomial_with_bare}, we discuss the relevant solutions for monomial couplings without and with a bare scalar potential, respectively.  For the busy reader, we summarise our main results in Sec.~\ref{sec:Summary}. We conclude with an extended discussion in Sec.~\ref{sec:Discussion}. We present many details of the calculations in the appendices. In particular, we study the case of a dilatonic coupling with bare mass in App.~\ref{app:dilaton_scalings}. Throughout the paper, we use units where $c=\hbar=1$ and $M_{\rm pl}^2 = 1/8\pi G$.

\section{Scalar-fermion Yukawa coupling}\label{sec:scalar-fermion}

We start by presenting the general equations and showing that the system generally contains a scaling regime due to an approximate scale invariance.

\subsection{Model and background equations}
We consider a dark sector composed of a scalar field $\varphi$ and a Dirac fermion $\psi$ described by the following action,
\begin{align}
S_{\rm DM}
=
\int \dd^4x\,&\sqrt{-g}\Big[-\frac{1}{2}g^{\mu\nu}\partial_\mu\varphi\,\partial_\nu\varphi
- V_{\rm bare}(\varphi)
\notag\\
&\,\,+\bar\psi\!\left(i\Gamma^\mu\nabla_\mu-m_\psi\right)\!\psi-y f(\varphi)\bar{\psi}\psi
\Big].
\label{eq:SDM_action}
\end{align}
Here $\Gamma^\mu$ are the curved-space gamma matrices, $m_\psi$ is the bare fermion mass, $V_{\rm bare}(\varphi)$ is a general bare scalar field potential to be fixed later. The scalar is coupled to fermions through a Yukawa-type interaction, 
\begin{equation}
    \mathcal{L}_\text{int} = -y f(\varphi)\bar{\psi}\psi\,,
\end{equation}
encoded in the function $f(\varphi)$, which we take to be of mass dimension 1. The parameter $y$ is a dimensionless constant corresponding to the standard Yukawa coupling. Throughout this work, we assume $y>0$ without loss of generality, as we leave the possibility of changing the sign of the interaction by changing the sign of $f(\varphi)$. Note that one may always use a field redefinition (see next subsection) to bring the interaction term to the standard Yukawa interaction. However, such a field redefinition would then have a non-canonical kinetic term. Unless otherwise stated, we work with the canonical form of the kinetic term.

One may wonder whether it is reasonable to consider such a decoupled dark sector. A plausible set-up is that the scalar-fermion sector belongs to a hidden dark sector that is populated after inflation, either directly through inflaton decay, gravitational particle production, or through a feeble portal interaction with the visible sector \cite{Hall:2009bx,Berlin:2016gtr,Berlin:2016vnh,Heurtier:2019beu,Garcia:2020eof,Brax_2022}. If the interactions between this sector and the Standard Model are sufficiently weak, the hidden sector decouples early from the thermal bath and subsequently evolves as an isolated subsystem. In contrast, the scalar and fermions may still interact efficiently with each other through the Yukawa coupling, so that energy can be exchanged internally while the total dark-sector energy is diluted only by the cosmological expansion. Fermion number is then conserved.

In this work, we mainly focus on a general monomial coupling given by
\begin{equation}
    f(\varphi)=\pm M_{\rm pl}\left(\frac{\varphi}{M_{\rm pl}}\right)^p\,,
\label{eq:fvarphi}
\end{equation}

with $p\geq1$ positive integer, although we briefly explore the general properties of the dilatonic coupling without bare mass of Ref.~\cite{Amendola:2017xhl} in Sec. \ref{eq:dilatoniclike} and explicit solutions with bare mass in App.~\ref{app:dilaton_scalings}.
For the monomial coupling \eqref{eq:fvarphi}, the effective coupling in the action \eqref{eq:SDM_action} is parametrised by $y_p\equiv y/M_{\rm pl}^{p-1}$ and has mass dimension $1-p$.
Thus, for $p>1$, the Yukawa-type operator $\mathcal{O}_p= y_p\varphi^p \bar\psi\psi$ is non-renormalisable and should only be interpreted as a higher-order effective field theory (EFT) operator of mass dimension $3+p$. Note that in the EFT approach, one should replace $M_{\rm pl}$ in Eq.~\eqref{eq:fvarphi} with a high-energy, ultraviolet (UV) cutoff scale, say $\Lambda_{\rm UV}$. For example, such couplings could appear after integrating out heavy fields with mass $M$.  In that case, the cut-off scale is $\Lambda_{\rm UV}=M$. One then expects
$y$ to be a dimensionless Wilson coefficient, expected to be of order unity. 

Note that, although a concrete construction is beyond the scope of this paper, one needs fine-tuning or symmetries to motivate large values of $p$ within the EFT approach. In the present analysis, we remain agnostic about the origin of the interactions and focus on their cosmological phenomenology. We note, however, that the EFT argument implies that larger values of $p$ should, in principle, lead to a stronger suppression of the coupling, provided the UV scale is fixed.

For the scalar field bare potential, we also consider a monomial ansatz, consistent with the coupling function $f(\varphi)$ \eqref{eq:fvarphi}. Namely, we assume that the bare potential in Eq.~\eqref{eq:SDM_action} is of the form
\begin{equation}
    V_{\rm bare}(\varphi)=\frac{\lambda_q}{q}M_{\rm pl}^4\,\left(\frac{\varphi}{M_{\rm pl}}\right)^q\,,
\end{equation}
where $q$ is constant, and we normalised the field to the Planck scale, for simplicity. For concrete cosmological applications, we focus on the cases where $q=\{2,4,6\}$.

\subsection{Equations of motion in the fluid approximation}

When working in the cosmological setup, we treat the fermions using the perfect-fluid approximation within the Kinetic gas approach. Namely, we consider that their energy-momentum tensor is well approximated by
\begin{align}
T_\psi^{\mu\nu}=(\rho_\psi+p_\psi)u^\mu u^\nu + p_\psi g_{\mu\nu}\,,
\label{eq:tmunufluid}
\end{align}
where $\rho_\psi$, $p_\psi$ and $u^\mu$ are respectively the energy density, pressure and 4-velocity of the fluid.  Their explicit expression for a degenerate gas can be found in App.~\ref{app:fermi_gas} and in general, in Ref.~\cite{Dom_nech_2021}. This approximation is valid as long as there are many fermions per Hubble volume.
From the second line of Eq.~\eqref{eq:SDM_action}, we see that the fermions acquire an effective mass given by
\begin{equation}
    m_{\rm eff}(\varphi)=\bigl|m_\psi+y f(\varphi)\bigr|\,,
    \label{eq:meff_def}
\end{equation}
where we take the absolute value to ensure that $m_{\rm eff}$ remains non-negative in the fluid description.

Next, we consider an FLRW spacetime in which the expansion is driven by a dominant cosmic fluid with constant equation-of-state parameter $w$. The metric in that case, and in cosmic time, reads
\begin{align}
\dd s^2=-\dd t^2+a^2(t)\dd x^2\,,\quad a(t)=a_i (t/t_i)^{2/(3+3w)}\,,
\label{eq:generalFLRW}
\end{align}
where $a(t)$ is the so-called scale factor and $t_i$ is an arbitrary pivot time, which we take as the initial time. From now on, all quantities with a subscript ``$i$'' are evaluated at $t_i$. In what follows, we set $a_i\equiv1$ for simplicity.  
We first keep $w$ arbitrary and later specialise to matter domination, radiation domination, and stiff-fluid (kination) domination, corresponding to $w=0$, $w=1/3$, and $w=1$, respectively. 

Notice that we assume that the scalar-fermion sector is subdominant and evolves on such a fixed cosmological background. This approximation is valid as long as its fractional energy density remains subdominant. In particular, for a stiff fluid background with $w=1$, the dark-sector energy density can redshift more slowly than the background and may eventually dominate, so the spectator approximation must be checked explicitly.

Varying the action \eqref{eq:SDM_action} with respect to $\varphi$ yields
\begin{equation}
    \ddot{\varphi}+3H\dot{\varphi}+\frac{\dd V_{\rm bare}(\varphi)}{\dd\varphi}
    +\frac{\partial \rho_\psi}{\partial\varphi}=0\,,
    \label{eq:scalar_eom_rho}
\end{equation}
where the last term is evaluated at a fixed fermion number, which is the relevant condition for the zero-temperature degenerate Fermi gas considered in this work. In that case, one then finds that
\begin{equation}
    \frac{\partial \rho_\psi}{\partial\varphi}
    =
    + y\,f_{,\varphi}(\varphi)\,
    \sgn\!\bigl(m_\psi+y f(\varphi)\bigr)\,
    \frac{\rho_\psi-3p_\psi}{m_{\rm eff}}\,,
    \label{eq:drhodphi_general}
\end{equation}
where $\sgn\!\bigl(m_\psi+y f(\varphi)\bigr)=\pm 1$ depending on the sign of the argument. 
For a general derivation in the case of fermions in a thermal bath, see Ref.~\cite{Dom_nech_2021}. Using Eq.~\eqref{eq:drhodphi_general}, the full scalar equation of motion \eqref{eq:scalar_eom_rho} reads
\begin{align}
    &\ddot{\varphi}+3H\dot{\varphi}+\frac{\dd V_{\rm bare}(\varphi)}{\dd\varphi}\notag\\
    &\quad +y\,f_{,\varphi}(\varphi)\,
    \sgn\!\bigl(m_\psi+y f(\varphi)\bigr)\,
    \frac{\rho_\psi-3p_\psi}{m_{\rm eff}}
    =0.
    \label{eq:scalar_eom_general}
\end{align}
In passing, we note that Eq.~\eqref{eq:drhodphi_general} is consistent with the formulation of Refs.~\cite{Amendola:2017xhl,Savastano:2019zpr}, since
\begin{align}
\frac{\partial \rho_\psi}{\partial\varphi}=\frac{\partial\ln m_{\rm eff}}{\partial\varphi}(\rho_\psi-3p_\psi)=-\frac{\partial\ln m_{\rm eff}}{\partial\varphi}\,T_\psi
\end{align}
where we used that $T_\psi=g_{\mu\nu}T_\psi^{\mu\nu}=-(\rho_\psi-3p_\psi)$.

Since the scalar-fermion system is totally decoupled from the background cosmic fluid, the dark sector, which we denote as ``DS'' henceforth, satisfies the continuity equation, 
\begin{equation}
    \dot\rho_\text{DS} + 3H(\rho_\text{DS} + p_\text{DS}) = 0\,,
    \label{eqn:DM_conservation}
\end{equation}
with $\rho_\text{DS} = \rho_\varphi + \rho_\psi$ and $p_\text{DS} = p_\varphi + p_\psi$ respectively. However, the energy exchange happens through the coupling, and it is manifested at the level of the continuity equation of each fluid, with the fermion fluid satisfying
\begin{equation}
    \dot{\rho}_\psi+3H(\rho_\psi+p_\psi)
    +\dot{\varphi}\,\frac{\partial \rho_\psi}{\partial\varphi}=0\,.
    \label{eq:fermion_conservation}
\end{equation}
Similarly, one can check that the scalar energy density satisfies
\begin{equation}
    \dot{\rho}_\varphi+3H(\rho_\varphi+p_\varphi)
    -\dot{\varphi}\,\frac{\partial \rho_\psi}{\partial\varphi}=0\,,
    \label{eq:scalar_conservation}
\end{equation}
where we define\footnote{Later in the paper we relabel the scalar density as $\rho_\phi$ as we work within the dimensionless field $\phi$ instead of the dimensionful $\varphi$.}
\begin{align}\label{eq:rhovarphidefinition}
\rho_
\varphi=\frac{1}{2}\dot\varphi^2+V_{\rm bare}(\varphi)\,.
\end{align}
By adding Eqs.~\eqref{eq:fermion_conservation} and \eqref{eq:scalar_conservation}, one confirms Eq. \eqref{eqn:DM_conservation}.

Eqs.~\eqref{eq:scalar_eom_general}
and~\eqref{eq:fermion_conservation} form the basic system for the coupled scalar-fermion sector. From the scalar field equations of motion \eqref{eq:scalar_eom_general}, we see that the coupling introduces a source term to the standard Klein-Gordon equation, whose shape determines the evolution of the system. Note that, while the total energy of the isolated scalar-fermion subsystem is still diluted by the cosmological expansion, how this energy is partitioned between the two sectors is controlled by the type of coupling.

\subsection{General argument for scaling solutions}
\label{subsec:conformal_intuition}

Before we examine specific solutions, it is instructive to study the theory's general asymptotic properties. In particular, in the absence of bare potential, that is $V_{\rm bare}=0$, Refs.~\cite{Dom_nech_2021,Amendola:2017xhl} find that the scalar-fermion system reaches a scaling solution in radiation domination for $m_{\rm eff}=|m_\psi + y\varphi|$ and $m_{\rm eff}=m_\psi e^{y\varphi/M_{\rm pl}}$, where all energy densities (scalar, fermion and background fields) scale like radiation, that is $\rho\propto a^{-4}$, regardless of whether the fermions are non-relativistic. Furthermore, they find that $m_{\rm eff}\sim 1/a$ in the scaling regime in both cases. However, it is unclear how general the scaling regime is and whether the $m_{\rm eff}\sim 1/a$ asymptotic behaviour is universal. Here, we provide a general argument for the scaling properties and derive the precise scale factor dependence of $m_{\rm eff}$. We later confirm our expectations with explicit examples. Note that one may naively argue that, since the fermions are relativistic for $m_{\rm eff}=0$, the system dynamically approaches a conformally invariant state, and hence all energy densities scale as radiation. As we show below, this is not always the case.

To understand the behaviour of the system near the $m_{\rm eff}\sim 0$ regime, let us look at the scalar-fermion action \eqref{eq:SDM_action} with a general scalar-dependent fermion mass function, say $m_{\rm eff}=m(\varphi)$, and set $V_{\rm bare}(\varphi)=0$ for simplicity. It is most convenient to focus first on $m(\varphi)$ and, from that, infer the scaling properties of the different fluids. Hence, we use $m(\varphi)$ as our scalar field and re-write \eqref{eq:SDM_action} as 
\begin{align}
S=\int \dd^4x\,\sqrt{-g}\Big[
-\frac12&K^2(m)g^{\mu\nu}\partial_\mu m\,\partial_\nu m
\nonumber\\&+i\bar\psi\Gamma^\mu\nabla_\mu\psi
-m\,\bar\psi\psi
\Big]\,,
\label{eq:conf_action_start}
\end{align}
where
\begin{align}
K(m)\equiv \left|\frac{\partial\varphi}{\partial m(\varphi)}\right|\,,
\end{align}  
with the understanding that, after inversion, $\varphi=\varphi(m)$. 

We now perform a conformal transformation of the metric, $ g_{\mu\nu}=\Omega^2 \tilde g_{\mu\nu}$, and a rescaling of the fermion field, namely $\tilde\psi=\Omega^{3/2}\psi$. The resulting action is given by
\begin{align}
S=\int d^4x\,\sqrt{-\tilde g}\Big[
-\frac12&\Omega^2K^2(m)\tilde g^{\mu\nu}\partial_\mu m\,\partial_\nu m
\nonumber\\&+i\bar{\tilde\psi}\tilde\Gamma^\mu\tilde\nabla_\mu\tilde\psi
-\Omega m\,\bar{\tilde\psi}\tilde\psi
\Big]\,.
\label{eq:conf_action_2}
\end{align}
We see that the action \eqref{eq:conf_action_start} is scale invariant only if $K(m)={\rm constant}$ as one can define $\tilde m=\Omega m$ in Eq.~\eqref{eq:conf_action_2} and recover \eqref{eq:conf_action_start}. For this reason, let us first consider the $K(m)\approx {\rm constant}$ case and later turn to a general $K(m)$.

\subsubsection{Yukawa-like interaction \label{sec:yukawageneral}}

The $K(m)={\rm constant}$ case exactly corresponds to $m=|m_\psi + y\varphi|$ and, therefore, to the standard Yukawa interaction studied in Ref.~\cite{Dom_nech_2021}. Note that in the absence of scalar potential, one can shift $\varphi= \tilde \varphi -m_\psi/y$ and set $m_\psi=0$ and $m=|y\tilde\varphi|$. This property also allows us to generalise the scaling regime to the monomial case $m=|m_\psi - yM_{\rm pl}(\varphi/M_{\rm pl})^p|$, where we chose the $-$ sign in $f(\varphi)$ \eqref{eq:fvarphi} to allow for $m=0$ for general $p$. After shifting $\varphi=\tilde \varphi +M_{\rm pl}(m_\psi/(yM_{\rm pl}))^{1/p}$, we have at leading order in $\tilde\varphi$ that $m\approx |y\tilde\varphi+...|$, since $\tilde\varphi\sim 0$ when $m\sim 0$. This means that the scalar-fermion system is approximately scale invariant for $f(\varphi)\propto \varphi^p$ near the $m\sim 0$ regime. One can also confirm this with an explicit calculation. Taking $m=|m_\psi-yM_{\rm pl}(\varphi/M_{\rm pl})^p|$, as in Eq.~\eqref{eq:fvarphi}, we have that
\begin{align}
K(m)=\frac{1}{yp}\left(\frac{yM_{\rm pl}}{m_\psi-m}\right)^{1-\frac{1}{p}}\,,
\end{align}
and, therefore, $K(m)\approx {\rm constant}$ when $m\ll m_\psi$.

To understand the scale-factor dependence of the solutions for $m$, note that, although the action \eqref{eq:conf_action_start} is not conformal invariant for $K(m)\approx {\rm constant}$, it can be approximately so if $|\nabla \Omega |\ll \Omega$. In this way, we can neglect derivatives of the conformal factor, and the scale symmetry becomes approximately conformal. This is precisely the case in an expanding background, where $\Omega = a(\tau)$ and $\tilde g_{\mu\nu}=\eta_{\mu\nu}$, with $\tau$ being the conformal time (that is, $dt=ad\tau$). In that case, $d\ln a/d\tau\propto 1/\tau$ and vanishes asymptotically. The system then dynamically approaches an approximate ``conformal invariant'' state. Neglecting derivatives of $a(\tau)$, the redefined field $\tilde m=am$ effectively lives in flat spacetime.

Note that we have so far ignored in the discussion a possible, intrinsic $m$-dependence inside $\bar\psi\psi$ in Eq.~\eqref{eq:conf_action_start}. This is consistent with non-relativistic fermions, as one expects that $\bar\psi\psi\sim n_\psi$, but not with relativistic fermions. To see this, note that the trace of the fermion energy momentum tensor evaluated on-shell gives
\begin{align}
T_\psi\big|_{\rm on-shell}=m_{\rm eff}\bar\psi\psi\,.
\end{align}
Using that in the fluid picture $T_\psi=-(\rho_\psi-3p_\psi)$, see Eq.~\eqref{eq:tmunufluid}, we find that
\begin{align}
\langle \bar\psi\psi\rangle =-\frac{\rho_\psi-3p_\psi}{m_{\rm eff}}\,.
\end{align}
Using the results of Sec.~\ref{sec:Fermi}, we find, in cosmology and for a degenerate Fermi gas, that   
\begin{align}
\label{eq:dependenceofpsipsi}
\langle \bar\psi\psi\rangle_{\rm non-rel} \propto a^{-3}\quad{\rm and}\quad \langle \bar\psi\psi\rangle_{\rm rel} \propto m a^{-2}\,,
\end{align}
where the brackets denote the expectation value of the fermion fluid.
This implies that in a cosmological background, the term $m\Omega\bar{\tilde\psi}\tilde\psi$ in Eq.~\eqref{eq:conf_action_2} is proportional to $am$ and $(am)^2$ respectively for non-relativistic and relativistic fermions. Thus, the scaling argument applies regardless of whether the fermions are relativistic.

From the $m$-dependence of Eq.~\eqref{eq:dependenceofpsipsi}, and as we show later in detail, the equations of motion for $\tilde m$ are approximately those of a harmonic oscillator. This means that $\tilde m$ oscillates around $\tilde m\sim 0$, and, therefore, it implies that $m\propto 1/ a \times {\rm oscillations}$. This is in agreement with the results of Ref.~\cite{Dom_nech_2021}. Note, however, that here we generalised this property to couplings of the type $f(\varphi)\propto \varphi^p$ with $p>1$ for the first time.

Regarding the energy density of the scalar field \eqref{eq:rhovarphidefinition}, we find, in the absence of bare potential and for the monomial $f(\varphi)$, that 
\begin{align}\label{eq:rhophirough}
\rho_\varphi& \propto \dot\varphi^2=\frac{ K^2(m)}{a^2}\left(\frac{\dd m}{a\dd\tau}\right)^2\sim\frac{1}{a^4}\times {\rm oscillations}\,,
\end{align}
where we used that $K\sim {\rm constant}$ and ${dm}/{d\tau}\sim 1/a \times {\rm oscillations}+{\cal O}(H/\omega)$ for oscillations with frequency $\omega \gg H$. From Eq.~\eqref{eq:rhophirough}, we see that the scalar field decays as radiation, regardless of whether fermions are relativistic or not. 

The energy density of the fermions also scales as radiation in the $m\sim 0$ regime. In the relativistic case, we trivially have $\rho_\psi\sim 1/a^4$. In the non-relativistic case, where $\rho_\psi\sim m n_\psi$, we have that $\rho_\psi \sim 1/a^4\times {\rm oscillations}$, since $m\sim 1/a$ and number density conservation implies $n_\psi\sim 1/a^3$. Thus, for a Yukawa-like interaction, we expect that $\rho_\varphi/\rho_\psi\approx {\rm constant}$ in the $m\sim 0$ regime, with both energy densities redshifting like a radiation fluid.

\subsubsection{Dilatonic-like interaction \label{eq:dilatoniclike}}

For a general $K(m)$, there are no such guiding symmetries. Nevertheless, one may generally expand $K(m)$ near the massless regime as $K(m)\sim m^\alpha$, assuming the relation $\varphi(m)$ allows it. For example, in the run-away mass dilatonic coupling of Ref.~\cite{Amendola:2017xhl}, namely $m(\varphi)=m_\psi e^{y\varphi/M_{\rm pl}}$, one has $K(m)={M_{\rm pl}}/({y\,m})$. For $K(m)\sim m^\alpha$, the action \eqref{eq:conf_action_2} is approximately given by
\begin{align}
S\approx \int \dd^4x\,\sqrt{-\tilde g}\Big[
-\frac12&\Omega^2m^{2\alpha}\tilde g^{\mu\nu}\partial_\mu m\,\partial_\nu  m
\nonumber\\&+i\bar{\tilde\psi}\tilde\Gamma^\mu\tilde\nabla_\mu\tilde\psi
-\Omega m\,\bar{\tilde\psi}\tilde\psi
\Big]\,.
\label{eq:conf_action_3}
\end{align}
Interestingly, in an expanding background and in conformal time $d\tau=dt/a$, we may rewrite Eq.~\eqref{eq:conf_action_3} as
\begin{align}
S\approx \int d^3xd\tau\,\Big[
\frac12&\left(a{\cal H}m^{1+\alpha}\right)^2\left(\frac{d\ln m}{d\ln a}\right)^2
\nonumber\\&+i\bar{\tilde\psi}\tilde\Gamma^\mu\tilde\nabla_\mu\tilde\psi
-a m\,\bar{\tilde\psi}\tilde\psi
\Big]\,,
\label{eq:conf_action_4}
\end{align}
where ${\cal H}=\tfrac{1}{a}\tfrac{da}{d\tau}=aH$.
We checked that $m$-field equations of motion allow for solutions of the type $m \propto a^{-\#}$, whenever the scale factor appears through a global prefactor in the action \eqref{eq:conf_action_4}, namely when  $(a{\cal H}m^{1+\alpha})^2=a m$ and $(a{\cal H}m^{1+\alpha})^2=(a m)^2$ for the non-relativistic and relativistic fermions, respectively. Note that, for consistency, ${\cal H}$ must also be a power-law of $a$.

For a cosmological background with a constant equation of state $w$, we have that ${\cal H}\sim a^{-(1+3w)/2}$. Then, the condition for the existence of power-law solutions in Eq.~\eqref{eq:conf_action_4} respectively reads
\begin{align}\label{eq:m_a_dependence}
m\propto a^{\frac{3w}{1+2\alpha}}\quad {\rm and}\quad m\propto a^{\frac{1+3w}{2\alpha}}\,,
\end{align}
for the non-relativistic and relativistic fermions. With our general results \eqref{eq:m_a_dependence}, we can check the runaway dilatonic mass case in radiation domination of Ref.~\cite{Amendola:2017xhl}, that is $\alpha=-1$ and $w=1/3$. There, we see that regardless of whether the fermions are relativistic, we have $ m \propto 1/a$. However, our general results show that the $m\propto 1/a$ scaling in radiation domination is coincidental. We also note that the effective fermion mass \eqref{eq:m_a_dependence} grows for $(w>0,\alpha<-1/2)$ and $(w>-1/3,\alpha>0)$, respectively, in the non-relativistic and relativistic cases, signalling that the $m\to 0$ regime is not an attractor solution in those cases. Since we are interested in the Yukawa-like coupling, we leave a detailed study of these cases for future work.

Lastly, let us discuss the energy density of the scalar and fermions in this case. From Eq.~\eqref{eq:m_a_dependence}, we also see that the energy density of the scalar field \eqref{eq:rhovarphidefinition} now reads
\begin{align}
\rho_\varphi& \propto\dot\varphi^2=  H^2m^2K^2(m)\left(\frac{\dd\ln m}{\dd\ln a}\right)^2\sim H^2 m^{2(1+\alpha)}\,.
\end{align}
Thus, there is a tracking solution, namely $\rho_\varphi\propto H^2$, only when $\alpha=-1$. For general values of $\alpha$, one has that $\rho_\varphi$ is a general power-law of $a$. We do not pursue this general direction further as it is not very illuminating. More interestingly, though, we see that the energy density of non-relativistic fermions scales as $\rho_\psi \sim m n_\psi\sim a^{-3+3w/(1+2\alpha)}$. For $\alpha=-1$ the fermion energy density also tracks the background, namely $\rho_\psi\propto H^2$. This shows that for non-relativistic fermions and $\alpha=-1$, the full system reaches a tracking solution. For relativistic fermions, however, one always has $\rho_\psi\propto a^{-4}$. We provide more details on the runaway dilatonic mass coupling in App.~\ref{app:dilaton_scalings}. In particular, we find that memory of the initial conditions for $w=1$ and logarithmic secular growth for $w=0$.

\section{Background dynamics for a Degenerate Fermi gas}\label{sec:Fermi}

We now specialise to the case where the fermions behave as a degenerate Fermi gas at zero temperature for analytical simplicity. The fermion number is then conserved, and all states are filled up to the Fermi momentum $p_F=(3\pi^2 n_\psi)^{1/3}$ with $n_\psi$ the fermion number density. For more details, see App.~\ref{app:fermi_exact}. It is convenient to redefine the number density as $N_\psi\equiv 3\pi^2 n_\psi$,
so that $p_F=N_\psi^{1/3}$. Since the total fermion number is conserved, the number density redshifts only due to the expansion of the universe,
\begin{equation}
    N_\psi=N_{\psi,i}a^{-3}\,,
\end{equation}
and therefore we need only investigate the behaviour of the scalar. 

In what follows, we are going to work with dimensionless ratios as in \cite{Dom_nech_2021}. First, to characterise the fermion gas, we introduce the dimensionless ratio
\begin{equation}
    x\equiv \frac{m_{\rm eff}}{N_\psi^{1/3}}\,,
\end{equation}
which measures the effective fermion mass relative to the Fermi energy. The limits where $x\ll1$ and $x\gg1$ correspond to \textit{relativistic} and \textit{non-relativistic} fermions, respectively. This will be a key quantity to characterise our solutions. 

The exact energy density and pressure of the degenerate Fermi gas are given by
\begin{equation}
    \rho_{\psi} = \frac{N_{\psi}^{4/3}}{8\pi^2}\,F(x),
    \qquad
    p_{\psi} = \frac{N_{\psi}^{4/3}}{24\pi^2}\,P(x),
    \label{eq:rho_p_exact_compact}
\end{equation}
with
\begin{align}
    F(x)&\equiv
    (2+x^2)\sqrt{1+x^2}
    -x^4\sinh^{-1}\!\left(\frac{1}{x}\right)\,,
    \\
    P(x)&\equiv
    (2-3x^2)\sqrt{1+x^2}
    +3x^4\sinh^{-1}\!\left(\frac{1}{x}\right)\,.
\end{align}
The Klein-Gordon equation \eqref{eq:scalar_eom_rho} for the scalar field is sourced by the trace of the fermion stress tensor, which we define as
\begin{equation}
    \mathcal S_\varphi
    \equiv
    \frac{\partial \rho_\psi}{\partial\varphi}
    =
    y\,f_{,\varphi}(\varphi)\,
    \sgn\!\bigl(m_\psi+y f(\varphi)\bigr)\,
    \frac{\rho_\psi-3p_\psi}{m_{\rm eff}}.
    \label{eq:source_def}
\end{equation}
From Eq.~\eqref{eq:rho_p_exact_compact}, a useful exact expression is given by
\begin{equation}
    \rho_\psi-3p_\psi
    =
    \frac{N_\psi^{4/3}}{2\pi^2}
    \left[
    x^2\sqrt{1+x^2}
    -x^4\arsinh\!\left(\frac{1}{x}\right)
    \right].
    \label{eq:rho_minus_3p_exact}
\end{equation}

The asymptotic limits of Eq.~\eqref{eq:rho_minus_3p_exact} are given by
\begin{equation}
    \rho_\psi-3p_\psi
    \simeq
    \frac{N_\psi^{4/3}}{2\pi^2}\left[x^2+\mathcal{O}\!\left(x^4\ln x\right)\right],
    \qquad x\ll1,
\end{equation}
and
\begin{equation}
    \rho_\psi-3p_\psi
    \simeq
    \frac{N_\psi^{4/3}}{3\pi^2}
    \left[
    x+\mathcal{O}(x^{-1})
    \right],
    \qquad x\gg1\,,
\end{equation}
respectively corresponding to the relativistic and non-relativistic regimes.

It is also convenient to define the initial fermion abundance given by
\begin{equation}
    \Omega_{\psi,i}\equiv \frac{\rho_{\psi,i}}{3M_{\rm pl}^2 H_i^2}.
\end{equation}
Using Eq.~\eqref{eq:rho_p_exact_compact}, one can express the initial fermion number density as
\begin{equation}
    N_{\psi,i}^{4/3}
    =
    \frac{24\pi^2 M_{\rm pl}^2 H_i^2}{F(x_i)}\,\Omega_{\psi,i},
    \label{eq:Ni_from_Omega}
\end{equation}
where
\begin{equation}
\label{eq:xi}
    x_i\equiv \frac{m_{{\rm eff},i}}{N_{\psi,i}^{1/3}}
    =
    \frac{|m_\psi+ y f(\varphi_i)|}{N_{\psi,i}^{1/3}}.
\end{equation}
These relations allow us to rewrite the equation of motion in terms of dimensionless cosmological quantities.  Detailed derivations of the formulas are collected in App.~\ref{app:fermi_gas}.

\subsection{Background equations of motion for monomial couplings}
\label{sec:representative_couplings}

Below we present the equations of motion for the scalar-fermion system for the monomial coupling \eqref{eq:fvarphi} written in dimensionless variables. This formulation is useful both numerically and for understanding the relativistic and non-relativistic limits analytically. A detailed discussion on the case of a dilatonic coupling with bare mass is given in App.~\ref{app:dilaton_scalings}. The monomial coupling provides a simple algebraic dependence of the effective mass on the scalar, while the exponential coupling can induce sharper variations. Let us remind the reader that we are working with a general FLRW metric with a constant equation of state given by Eq.~\eqref{eq:generalFLRW}

From Eq.~\eqref{eq:SDM_action} and \eqref{eq:fvarphi}, we see that by introducing the dimensionless coupling strength given by
\begin{equation}
    \beta \equiv \left(\frac{y M_{\rm pl}}{m_\psi}\right)^{1/p},
\end{equation}
where $\beta>0$ since we assumed $y>0$, we can define the dimensionless scalar field as
\begin{equation}\label{eq:dimensionlessphi}
    \phi \equiv \frac{\beta\varphi}{M_{\rm pl}}\,.
\end{equation}
This expression is analogous to that of Ref.~\cite{Dom_nech_2021}, as for $p=1$ we have $\beta=yM_{\rm pl}/m_\psi$, which reduces to the usual linear-coupling parameter. Using Eq.~\eqref{eq:dimensionlessphi},
the effective mass reads $m_{\rm eff}=m_\psi |1\pm \phi^p|$.  Also, from the definition of the initial ratio $x_i$ \eqref{eq:xi},  it follows that
\begin{equation}
    x=\frac{m_{\rm eff}}{N_\psi^{1/3}}
    =
    x_ia
    \frac{|1\pm \phi^p|}{|1\pm \phi_i^p|}\,.
    \label{eq:x_monomial}
\end{equation}

Note that the scalar-fermion system may have solutions for which $m_\text{eff}=0$. Since we find that the scalar field oscillates around $m_\text{eff}=0$ (see also the general discussion in Sec.~\ref{sec:yukawageneral}), we denote the regime when $m_\text{eff}\sim0$ as ``saturation point''. The existence of such a saturation point depends both on the sign of the coupling and on the parity of $p$. In terms of the dimensionless field, we have that for $f(\phi)\propto+\phi^p$ the saturation requires $1+\phi^p=0$, which gives 
$\phi_s=-1$ for odd $p$ and no real saturation point for even $p$.  For $f(\phi)\propto-\phi^p$, saturation requires $1-\phi^p=0$, giving $\phi_s=+1$ and $\phi_s=\pm 1$ for odd and even $p$, respectively. 

Writing the scalar equation \eqref{eq:scalar_eom_general} in terms of derivatives with respect to the scale factor, we obtain
\begin{equation}
    \phi''+\frac{5-3w}{2a}\phi'
    +\Lambda_q a^{3w+1}\phi^{q-1}
    +\frac{\beta}{M_{\rm pl} a^2H^2}\mathcal S_\varphi=0,
    \label{eq:phi_monomial_general}
\end{equation}
where we defined $\phi'=d\phi/da$, and we introduced the bare potential contribution parameter defined by
\begin{equation}
    \Lambda_q\equiv \frac{\lambda_q M_{\rm pl}^{2}}{H_i^2\beta^{q-2}}\,.
\end{equation}
For later convenience, we separately show the limits of Eq.~\eqref{eq:phi_monomial_general} for relativistic and non-relativistic fermions below.

First, in the relativistic regime, that is $x\ll1$, Eq.~\eqref{eq:phi_monomial_general} reduces to
\begin{equation}
    \phi''+\frac{5-3w}{2a}\phi'
    +\Lambda_q a^{3w+1}\phi^{q-1}
    \pm Kpa^{3w-1}(1\pm \phi^p)\phi^{p-1}=0,
    \label{eq:phi_rel_main}
\end{equation}
where we defined
\begin{equation}
    K\equiv \frac{12\,\beta^2 x_i^2}{F(x_i)\,|1\pm\phi_i^p|^2}\,\Omega_{\psi,i}\,.
    \label{eq:K_def}
\end{equation}
Note that, for relativistic initial conditions, that is $x_i\ll1$, one has $F(x_i)\simeq 2$, so that
\begin{equation}
    K \simeq \frac{6\,\beta^2 x_i^2}{|1\pm\phi_i^p|^2}\,\Omega_{\psi,i}\,.
\end{equation}

Second, in the non-relativistic regime, that is $x\gg1$, the equation of motion \eqref{eq:phi_monomial_general} becomes
\begin{align}
    \phi''+\frac{5-3w}{2a}\phi'
    &+\Lambda_q a^{3w+1}\phi^{q-1}\nonumber\\&
    \pm Dpa^{3w-2}\sgn(1\pm \phi^p)\phi^{p-1}=0\,,
    \label{eq:phi_nr_main}
\end{align}
where we have now defined
\begin{equation}
    D\equiv \frac{8\,\beta^2 x_i}{F(x_i)\,|1\pm\phi_i^p|}\,\Omega_{\psi,i}\simeq \frac{3\,\beta^2}{|1\pm\phi_i^p|}\,\Omega_{\psi,i}\,,
    \label{eq:D_def}
\end{equation}
for \(x_i\gg1\), where \(F(x_i)\simeq \frac{8}{3}x_i\).

For all practical purposes, the source constants $K$ and $D$, respectively given by Eqs.~\eqref{eq:K_def} and \eqref{eq:D_def}, play a significant role in the evolution. For instance, one must specify the initial conditions for the scalar field, namely the initial $\phi(a_i)$ and $\phi'(a_i)$, as well as the initial fermion fractional density $\Omega_{\psi\,i}$ and the initial ratio $x_i$. Although there is ample freedom in their specification, we make the following simplifying assumptions when evolving the scalar equation of motion. First, we assume $\Omega_{\psi\,i},\Omega_{\psi\,i}
\ll 1$, as they should be small enough for the spectator approximation to be valid.  Second, we assume that $\phi_i$ is fixed away from the saturation point. More concretely, we require that initially $|1\pm\phi_i^p|\sim\mathcal{O}(1)$. The latter arbitrary condition is intended to prevent the field from getting stuck at $\phi\sim0$ from the start in the general case. Although this is a valid solution, it is not interesting for obvious reasons.

Regarding the energy density of the scalar field \eqref{eq:rhovarphidefinition}, we find that it is given by
\begin{equation}
    \rho_\phi
    \simeq
    \frac{M_{\rm pl}^2H_i^2}{\beta^2}
    \left[
    \frac12 a^{2-3(1+w)}\phi'^2 + \frac{\Lambda_q}{q}\phi^q
    \right]\,.
    \label{eq:rho_phi_bare}
\end{equation}
This allows us to readily compare the scalar energy density with that of the fermions. Expanding Eq. (\ref{eq:rho_p_exact_compact}), one finds,
in the relativistic regime where $F(x)\simeq F(x_i)\simeq2$, that
\begin{equation}
    \rho_\psi = 3M_\text{Pl}^2 H_i^2\Omega_{\psi\,,i}a^{-4}\,,
\end{equation}
which indeed confirms that the relativistic fermions decay as a radiation fluid. Then, the energy density ratio in the relativistic regime is given by
\begin{equation}
    \label{eq:rhopsinonrel}
    \left.\frac{\rho_\phi}{\rho_\psi}\right|_{\rm rel}
    \simeq
    \frac{1}{3\beta^2\Omega_{\psi,i}}
    \left[
    \frac12 a^{6-3(1+w)}\phi'(a)^2
    +
    \frac{\Lambda_q}{q}a^4\phi(a)^q
    \right]\,.
\end{equation}

On the other hand, in the non-relativistic regime, after using Eq. (\ref{eq:rho_p_exact_compact}), the fermion energy density reads
\begin{equation}
\label{eq:rhopsinonrel}
    \rho_\psi = \frac{3M_\text{Pl}^2H_i^2\Omega_{\psi,i}}{|1\pm\phi_i^p|}a^{-3}|1\pm\phi^p|\,.
\end{equation}
From Eq.~\eqref{eq:rhopsinonrel} we see that the fermion energy density seemingly decays as matter fluid due to the $a^{-3}$ factor as long as $\phi$ remains approximately constant. However, the story is different if they reach the scaling solution around the saturation point. In that case, the argument in the absolute value decays with the envelope $\sim a^{-1}$ and thus $\rho_\psi\sim a^{-4}$. See also the discussion in Sec.~\eqref{subsec:conformal_intuition}. From Eqs.~\eqref{eq:rho_phi_bare} and \eqref{eq:rhopsinonrel} we arrive at 
\begin{align}
    \left.\frac{\rho_\phi}{\rho_\psi}\right|_{\text{non-rel}}
    \simeq &
    \frac{|1\pm\phi_i^p|}{3\beta^2\Omega_{\psi,i}|1\pm\phi^p|}\nonumber\\&\times
    \Big[
    \frac12 a^{5-3(1+w)}\phi'(a)^2
    +
    \frac{\Lambda_q}{q}a^3\phi(a)^q
    \Big].
\end{align}
We proceed by investigating explicit solutions.

\section{Solutions for monomial couplings with $V_{\rm bare}=0$}
\label{sec:monomial_results}

We now focus on the scalar dynamics for monomial coupling \eqref{eq:fvarphi}. Recall that the scalar field acquires an effective potential induced by its coupling to the fermion fluid, and it evolves dynamically toward the relevant attractor of the system. This is clear if we define an effective dimensionless potential, consistent with Eq.~\eqref{eq:phi_monomial_general}, from the bare potential and the integration of the fermion source term as
\begin{align}
    V_{\rm eff}(\varphi\,;\,a)=\frac{V_{\rm bare}(\varphi)}{a^2H^2M^2_{\rm pl}}+\frac{1}{a^2 H^2M^2_{\rm pl}}\int \dd\varphi \,\mathcal{S}_{\varphi}\,,
    \label{eq:veff}
\end{align}
where $S_\varphi$ is given by Eq.~\eqref{eq:source_def},
and drop any constant term since it does not affect the dynamics of the scalar-fermion spectator system.  Note that the scale factor dependence appears directly from Eq.~\eqref{eq:phi_monomial_general}.\footnote{In practice, if we transform the equations back to the original $t$ variable, then the scalar equation of motion will consist of the $\dd V_\text{bare}/\dd\varphi$ term plus the fermion induced term, which decays and therefore at late times the effective potential is dominated by the bare term.} In this section, we first set the bare scalar potential to zero. We recover the bare potential later in Sec.~\ref{sec:monomial_with_bare}.

The late-time behaviour depends strongly on three ingredients: the parity of $p$, whether the fermions are relativistic or non-relativistic, and the background equation-of-state parameter $w$. Broadly speaking, two distinct types of evolution can occur:
\begin{itemize}
    \item An oscillatory branch, where the field settles near a minimum of the effective potential and undergoes damped oscillations. For the branches that approach the saturation point, the oscillation envelope scales as $a^{-1}$;
    \item An asymptotic branch, where the field approaches a flat region of the effective potential and decays mainly through Hubble friction. This branch appears for sufficiently flat potentials, in particular for $p>2$.
\end{itemize}
We proceed to study the relativistic and non-relativistic regimes separately below.

\subsection{Relativistic fermions and $f(\varphi)\propto +\varphi^p$}\label{subsec:relplusvzero}

In the relativistic limit, the solution of Eq.~\eqref{eq:phi_rel_main} may be viewed as the motion in the time-dependent effective potential \eqref{eq:veff} that reduces to
\begin{equation}
    V_{\rm eff}^{\rm rel}(\phi\,;\,a)
    =
    K
    \left(
    \phi^p+\frac12\phi^{2p}
    \right)a^{3w-1}\,.
    \label{eq:rel_general_veff}
\end{equation}
Note that its overall normalisation scales as $a^{3w-1}$. From this, we conclude that the effective potential decreases during matter domination ($w=0$), remains constant during radiation domination ($w=1/3$), and increases during stiff-fluid domination ($w>1/3$). The latter case is special because the fermion-induced contribution to the scalar dynamics becomes increasingly important. This is expected, since in a stiff-fluid background the dominant energy density redshifts as $a^{-6}$ compared to $a^{-4}$ and $a^{-3}$ in radiation and matter domination, respectively. 

For odd $p$, the potential has a minimum at $\phi_s=-1$. For $p=1$ in particular, this is the only stationary point and the field is always driven towards it, corresponding to $m_{\rm eff}\to0$. For $p>1$, the point $\phi=0$ also becomes stationary, although it is not a minimum but an inflexion point instead. If the field is initially placed exactly at $\phi_i=0$ with $\phi_i'=0$, in the absence of fluctuations, it remains there forever. In that special case, initially relativistic fermions with mass equal to their bare mass remain on that branch until the expansion makes them non-relativistic. However, any generic perturbation away from $\phi=0$ drives the field toward one of the dynamical branches described above. We first study the case with $p=1$ and later turn to the more general cases of odd $p>1$ and even $p$. At the end of this subsection, we discuss solutions for $p>2$ which oscillate around $\phi\sim 0$.

\subsubsection{The $p=1$ case}
For $p=1$, corresponding to the standard Yukawa coupling studied in Ref.~\cite{Dom_nech_2021}, the equation of motion \eqref{eq:phi_rel_main} becomes
\begin{equation}
    \phi''+\frac{5-3w}{2a}\phi'
    +Ka^{3w-1}(1+\phi)=0.
    \label{eq:linear_p1_rel}
\end{equation}
For $w\neq-1/3$, this equation has a closed-form solution in terms of Bessel functions,\footnote{Using the initial conditions of \cite{Dom_nech_2021}, that is that for $\phi_i = \phi_i' = 0$, we obtain $c_1 = - \pi \frac{2\sqrt{K}}{3w+1}Y_{\nu+1}\left(\frac{2\sqrt{K}}{3w+1}\right)$ and $c_2 = \pi \frac{2\sqrt{K}}{3w+1} J_{\nu+1}\left(\frac{2\sqrt{K}}{3w+1}\right)$.} namely
\begin{equation}\label{eq:phisolutionrel}
    \phi = -1 + a^{\frac{3(w-1)}{4}}\left(c_1J_\nu(z(a)) + c_2Y_\nu(z(a))\right),
\end{equation}
where
\begin{equation}
    \nu = \frac{3(1-w)}{2(3w+1)}
    \quad{\rm and}\quad 
    z(a) = \frac{2\sqrt{K}}{3w+1}a^{\frac{3w+1}{2}}.
\end{equation}
Note that, although we have checked that $z(a)$ is proportional to conformal time, that is $z(a)\propto \tau$ for any background $w$ parameter, we prefer to work with $a$ as we are interested in the scale factor dependence of the envelope of $\phi(a)$.

From the asymptotic limits of the Bessel functions, we find that the asymptotic behaviour of Eq.~\eqref{eq:phisolutionrel} at late times, that is $a\gg a_i=1$, is given by
\begin{equation}
     \phi(a)=
    -1+\frac{A}{a}
    \cos\left(
    \frac{2\sqrt{K}}{3w+1}a^{\frac{3w+1}{2}}+\theta
    \right).
\end{equation}
Here $A$ is an overall constant fixed by the initial conditions, and $\theta$ is an unimportant constant phase. Thus, the field oscillates around the saturation point with an envelope proportional to $a^{-1}$. For fixed initial conditions and fixed initial fermion number density, the scalar reaches the saturation point the fastest in a stiff-fluid background and slowest in a matter-dominated background. We find that this qualitative ordering also holds in the non-relativistic cases and for the other couplings considered below. In this sense, a stiff-fluid background drives the system toward the scaling solution most efficiently.

\subsubsection{Odd $p>1$ case}

For any odd $p>1$, the effective potential also has a minimum at $\phi_s=-1$. Linearising (\ref{eq:linear_p1_rel}),
we obtain, for $w\neq -1/3$, the late-time oscillatory solution around $\phi_s=-1$, which reads
\begin{equation}
    \phi(a)\approx
    -1+\frac{A_p}{a}
    \cos\left(
    \frac{2p\sqrt{K}}{3w+1}a^{\frac{3w+1}{2}}+\theta_p
    \right)\,.
    \label{eq:phi_odd_osc}
\end{equation}
Note that Eq.~\eqref{eq:phi_odd_osc} has the same envelope as the $p=1$ solution \eqref{eq:phisolutionrel}, with the frequency rescaled by a factor $p$. This asymptotic behaviour around the saturation point is consistent with the general arguments presented in Sec.~\ref{sec:yukawageneral}. Note that the constants $A_p$ and $\theta_p$ might differ from the case $p=1$ by order unity factors. 

\begin{figure}[t]
\includegraphics[width=0.5\textwidth]{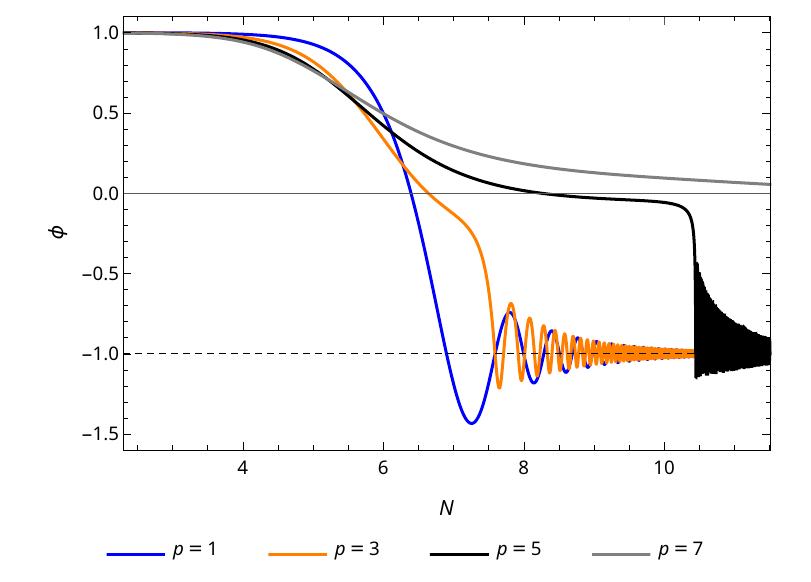}
\caption{The solution for $\phi$ as a function of the number of efolds for a radiation-dominated background, $w=1/3$, with monomial couplings $p=1$ (blue line), $p=3$ (orange line), $p=5$ (black line), and $p=7$ (grey line). We take $K=10^{-5}$, relativistic fermions, and scalar initial conditions $\phi_i=+1$ and $\phi_i'=0$. The scalar field is driven toward the saturation point $\phi_s=-1$ for $p=1$, $p=3$, and $p=5$. For $p=7$, and for higher powers with the same initial conditions, the field approaches $\phi=0$ asymptotically, and the fermions recover their bare mass. The late-time behaviour of the oscillatory solutions becomes harmonic, as expected from Eq.~\eqref{eq:phi_odd_osc}, see Sec. \ref{subsec:relplusvzero}.}
\label{fig:w1/3}
\end{figure}

Eq.~\eqref{eq:phi_odd_osc} is the late-time state whenever the field evolves into the basin of attraction of the minimum. The difference between different powers $p$ is that, for larger $p$, the effective potential becomes flatter around $\phi=0$. Thus, starting again from $\phi_i=+1$ with $\phi_i'=0$, the field needs more time to cross the flat region than in the $p=1$ case, where the effective potential has a quadratic form and the field is therefore directly driven toward $\phi_s=-1$. In Fig.~\ref{fig:w1/3} we present the numerical evolution of the scalar field in terms of efolds ($N=\ln a$) for different odd-power couplings with fixed initial conditions $\phi_i = 1$, $\phi_i'=0$ and source $K = 10^{-5}$ and for a radiation-dominated background. The key conclusion is that for lower $p$ values it is easier for the field to reach the saturation point at $\phi_s=-1$, while for larger $p$ the fact that the effective potential becomes flatter near $\phi=0$ makes the field slow down due to Hubble friction and possibly enter the asymptotic regime. Concretely, we find that after $p=7$ with these particular initial conditions, all radiation background solutions become asymptotic with $\phi\sim a^{-2/(p-2)}$. Notice also that initially the oscillations start increasingly non-symmetric as the effective potential becomes less symmetric around $\phi_s=-1$ for increasing value of $p$, although it eventually approaches the behaviour of Eq.~\eqref{eq:phi_odd_osc}.

\begin{figure}[t]
\includegraphics[width=0.5\textwidth]{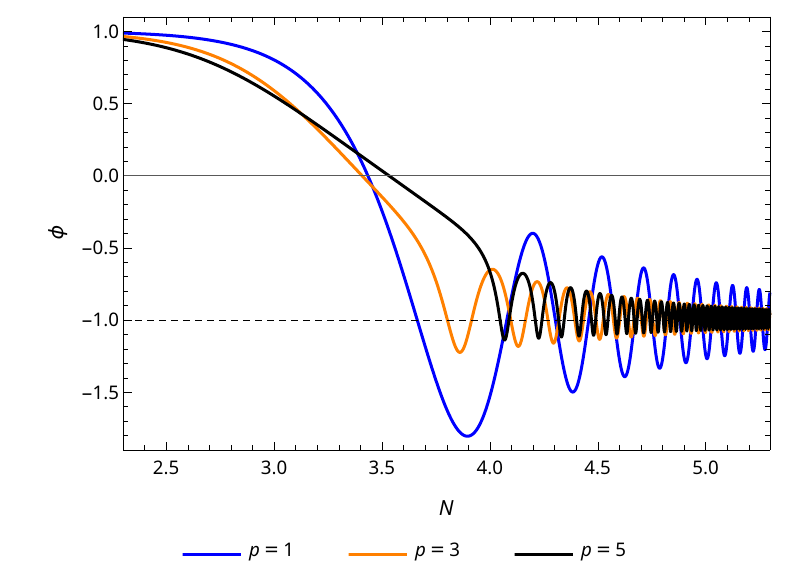}
\caption{The solution for $\phi$ as a function of the number of efolds for a stiff-fluid background, $w=1$, with monomial couplings $p=1$ (blue line), $p=3$ (orange line), and $p=5$ (black line). We take $K=10^{-5}$, relativistic fermions, and scalar initial conditions $\phi_i=+1$ and $\phi_i'=0$. The scalar field is driven toward the saturation point $\phi_s=-1$ in all cases shown, and no asymptotic solution has been found. Compared with the radiation background, the scalar field reaches saturation faster in a stiff fluid, see Sec. \ref{subsec:relplusvzero}.}
\label{fig:w1}
\end{figure}

In Fig. \ref{fig:w1}, we present the same system, but now in a stiff-fluid-dominated background with $w=1$. As expected, we find that the late-time behaviour is always oscillatory, as the effective potential becomes increasingly important and drags the scalar field towards the saturation point. The late time oscillations behave exactly as predicted from Eq. (\ref{eq:phi_odd_osc}). We also notice that the scalar field reaches the scaling regime much faster compared to the radiation background, acquiring larger kinetic energy compared to radiation and even more compared to the matter background. Consequently, the oscillation amplitude is generally larger for $w=1$ compared to the $w=1/3$ and even larger compared to matter.

Here we have to make some remarks regarding the numerical value of $K$ we choose. By definition, $K$ contains both $\Omega_{\psi\,i}$, $\beta$ and $x_i$, which gives us a lot of freedom for specifying their values. Nevertheless, one needs to be careful that first, the dark sector does not dominate sufficiently quickly in a stiff-fluid background, and at the same time $x$ does not grow such that the fermions become non-relativistic. For practical purposes, we have that $x_i\lesssim10^{-3}$, $\Omega_{\psi\,i}\lesssim10^{-3}$, $\beta\gtrsim10^2$ for radiation and for stiff fluid as the relativistic fermions decay slower than the stiff-fluid background, we need $\Omega_{\psi\,i}\lesssim10^{-5}$. Note that large values of $\beta$ only imply that the bare mass of the fermions is much smaller than the Yukawa coupling, and such a choice is not problematic from the EFT point of view as $m_\psi\ll M_{\rm pl}$. Moreover, the saturation region \(\phi=\mathcal O(1)\) corresponds to \(\varphi/M_{\rm pl}\sim 1/\beta\), so for \(\beta\gg1\) the dynamics takes place in a parametrically sub-Planckian field range.

Since the oscillation envelope on the saturation branch \eqref{eq:phi_odd_osc} scales as $a^{-1}$, it follows that the scalar energy density \eqref{eq:rhovarphidefinition} redshifts as $\rho_\phi\sim a^{-4}$, namely as a radiation fluid. The fermion fluid scales in the same way in the relativistic regime, by definition. Therefore, the saturation branch provides a radiation-like scaling: the scalar and fermion energy densities both redshift as radiation, and their ratio remains approximately constant. In fact, we find that at leading order
\begin{equation}
    \left.
    \frac{\langle\rho_\phi\rangle}{\langle\rho_\psi\rangle}
    \right|_{\rm rel}
    \simeq
    \frac{A_p^2 p^2 x_i^2}
    {2|1\pm \phi_i^p|^2}\,,
    \label{eq:rho_phi_over_rho_psi_rel_oddp}
\end{equation}
where $\langle...\rangle$ denotes oscillation average.
Additionally, the fact that $1+\phi\sim a^{-1}$ implies that $x$ defined in the previous section as the field enters the saturation regime, $\langle x\rangle\sim a\langle|1+\phi|\rangle\sim a^0$ and therefore approaches a constant value. Thus, relativistic fermions entering this regime will always remain relativistic regardless of the background cosmology. This does not apply to asymptotic solutions, or oscillatory solutions around $\phi=0$, as we show later. 

\subsubsection{The even $p$ case}

Let us now turn to oscillatory solutions with even $p$. In this case, the effective potential has a minimum at $\phi=0$. Thus, the effective mass cannot reach zero, and the field eventually stabilises at $\phi=0$, where the fermions recover their bare mass $m_\psi$. We start with the simplest case, $p=2$, and then turn to the general case.

The $p=2$ case is special because, after linearising the relativistic equation of motion around $\phi=0$, the source term becomes $Kpa^{3w-1}(1+ \phi^p)\phi^{p-1}\approx Kpa^{3w-1}\phi^{p-1}$. Thus, for $p=2$, the late-time behaviour is that of a harmonic oscillator. The obtained solution is then, 
\begin{equation}
    \phi(a)\approx\frac{A}{a}
    \cos\left(
    \frac{2\sqrt{2K}}{3w+1}a^{\frac{3w+1}{2}}+\theta
    \right),
    \label{eq:p2_rel_solution}
\end{equation}
again implying that the scalar energy density $\rho_\phi\sim a^{-4}$ at late times.

For $p>2$ even, the source term does not become linear near $\phi=0$, and therefore the oscillations around the origin are not expected to be purely harmonic. Numerically, in the cases where oscillations occur, we find an envelope consistent with $\phi=  a^{-4/(p+2)}\times\text{oscillations}(a)$. The oscillations also become less harmonic as $p$ increases.

\subsubsection{Solutions with $\phi\sim 0$ for $p>2$}

For $p>2$, \textit{both even and odd}, we also have found configurations with $\phi_i>0$ and $\phi_i'=0$ in which the field slows down as it approaches $\phi=0$ asymptotically\footnote{Whether the scalar field enters the asymptotic regime or oscillates around for $p$ even or goes towards the saturation regime at $\phi_s=-1$ is linked with scalar initial conditions.}. In this case, the fermion mass asymptotes to its bare value. We find that at leading order, when the field enters the asymptotic regime, it always behaves as
\begin{equation}
\phi(a) \sim
    a^{-\frac{3w+1}{p-2}}\,,
\end{equation}
for $w\leq1/3$. 
For $w=0$ and $w=1/3$, we obtain
\begin{equation}
    \frac{\rho_\phi}{\rho_\psi}\sim
    \begin{cases}
        a^{1-\frac{2}{p-2}}, & w=0,\\[1mm]
        a^{-\frac{4}{p-2}}, & w=\frac13.
    \end{cases}
    \label{eq:rel_ratio_scaling}
\end{equation}
Hence, in matter domination, the scalar can eventually dominate for sufficiently large $p$, while in radiation domination it always redshifts faster than the fermions on this non-oscillatory branch. In practice, for initial conditions $\phi_i = +1$ and $\phi_i'=0$, we find that the field enters this asymptotic regime for $p\geq5$ when $w=0$, and for $p\geq7$ when $w=1/3$. This creates the possibility that, if the field starts with inhomogeneities, different patches may enter different late-time branches: some may enter the asymptotic branch, while others may end up near the minimum of the effective potential.

Notice also that such an asymptotic solution does not exist for stiff fluids, as the effective potential becomes more important at late times. Unless the field starts exactly with $\phi_i=\phi_i'=0$, the potential becomes steeper with increasing scale factor, and the field eventually oscillates around its minimum, either $\phi=-1$ for odd $p$ or $\phi=0$ for even $p$. 

One may worry that, as the field asymptotically approaches zero, the fermions might transition from the relativistic to the non-relativistic regime, thereby modifying the evolution. However, once the field enters the asymptotic regime, the effective potential remains flat near $\phi=0$ also in the non-relativistic regime, as we discuss in the next subsection. We therefore expect that an asymptotic solution in the relativistic limit will transit to an asymptotic solution also in the non-relativistic limit.

\subsection{Non-relativistic fermions and $f(\varphi)\propto+\varphi^p$}\label{sec:nonrelplus}

In this section, we focus on the dynamics of the scalar field when fermions are non-relativistic.
In the non-relativistic limit, Eq.~\eqref{eq:phi_nr_main} may be interpreted as motion in the effective potential \eqref{eq:veff} that reduces to
\begin{equation}
    V_{\rm eff}^{\rm non-rel}(\phi;a)
    =
    D a^{3w-2}
    \begin{cases}
        \phi^p, & p \ {\rm even}\,,\\
        |1+\phi^p|, & p \ {\rm odd}\,.
    \end{cases}
    \label{eq:nr_veff_general}
\end{equation}
Here, the overall normalization scales as $a^{3w-2}$, unlike in the relativistic case \eqref{eq:rel_general_veff}, where the scaling is $a^{3w-1}$. Thus, the effective potential decays during both matter and radiation domination, and grows only in a stiff-fluid background with $w>2/3$.

\subsubsection{The odd $p$ case}

For odd $p$, the saturation point lies at the minimum of the effective potential at $\phi_s=-1$. To understand the local behaviour around this point, we define $u=\phi+1$ with $|u|\ll1$.
Expanding around $\phi_s=-1$, one finds
\begin{equation}
    1+\phi^p
    =
    p u-\frac{p(p-1)}{2}u^2+{\cal O}(u^3),
\end{equation}
and
\begin{equation}
    \phi^{p-1}
    =
    1-(p-1)u+{\cal O}(u^2).
\end{equation}
Therefore, close to the saturation point, the non-relativistic equation \eqref{eq:phi_nr_main} reduces to
\begin{equation}
    u''
    +
    \frac{5-3w}{2a}u'
    +
    Dp a^{3w-2}{\rm sgn}(u)
    =
    {\cal O}\left(a^{3w-2}|u|\right)\,.
    \label{eq:nr_oddp_cusp_eom_a}
\end{equation}
The leading term corresponds to a cusp-like local potential,
\begin{equation}
    V_{\rm eff}^{\rm non-rel}(u)
    \simeq
    Dp a^{3w-2}|u|.
\end{equation}
The force is therefore approximately constant on each side of the cusp. As a result, the oscillations around $\phi_s=-1$ are not harmonic. Instead, at leading order, the waveform is piecewise parabolic. The explicit parabolic-wave solution ansatz is given in Appendix~\ref{app:Parabolic_waves}; here we keep only the first harmonic of its Fourier expansions, which captures the envelope and the leading phase of the oscillations.

The resulting late-time solution may be approximated as
\begin{equation}
    \phi(a)
    \approx
    -1
    +
    \frac{32 A_p}{\pi^3 a}
    \cos\Big(
        \sqrt{\frac{\pi^2 Dp}{2(3w+1)^2 A_p}}
        a^{\frac{3w+1}{2}}
        + \Theta_p(a)
    \Big)\,.
    \label{eq:nr_oddp_first_harmonic}
\end{equation}
The matching constant $A_p$ depends on the initial conditions and on the transient evolution before the field settles into the saturation regime. The factor $32/\pi^3$ is the first Fourier coefficient of the unit-amplitude parabolic wave.
The leading phase scales as $a^{(3w+1)/2}$, which is identical to the phase dependence in the relativistic regime. 

When the field starts far from $\phi_s=-1$, the transient evolution through the non-asymptotic part of the potential can produce a noticeable phase shift. Numerically, this appears as a slow drift between Eq.~\eqref{eq:nr_oddp_first_harmonic} and the full solution, especially for a stiff fluid background. This drift can be captured by the subleading correction $\Theta_p(a)$, which in $w=1$ and $w=0$ backgrounds includes logarithmic terms in finite-time fits. These corrections affect the detailed matching of the zero crossings, but they do not modify the leading envelope, which remains the $a^{-1}$. We also notice that for $p=1$, the phase correction $\Theta_p(a)$ has very small $a$ dependence as the source term on the right-hand side of Eq. (\ref{eq:nr_oddp_cusp_eom_a}) has second-order dependence on the displacement from $\phi_s$. 

Specifically, we confirm numerically that, whenever the field reaches the saturation regime, the late-time decay envelope is robust, $|\phi+1|\sim a^{-1}$. This can be checked directly by extracting the peaks of the numerical solution. For larger odd $p$ and especially for $w=1$, the maxima and minima can initially have slightly different envelopes because the subleading terms in Eq.~\eqref{eq:nr_oddp_cusp_eom_a} are still relevant and also as the cusp minimum is asymmetric. After the field evolves sufficiently close to the saturation point, however, both envelopes converge to the same $a^{-1}$ scaling.

This scaling immediately implies that the non-relativistic fermion energy density on the saturation branch redshifts as radiation:
\begin{equation}
    \rho_\psi
    \propto
    a^{-3}|1+\phi^p|
    \simeq
    p a^{-3}|\phi+1|
    \propto
    a^{-4}.
\end{equation}
The scalar energy density has the same leading scaling after coarse-graining. Indeed, since the phase in Eq.~\eqref{eq:nr_oddp_first_harmonic} is rapidly varying, the dominant contribution to $\phi'$ comes from differentiating the dominant phase and thus, $\rho_\phi\sim a^{-4}$ similarly to the relativistic case, up to subleading logarithmic corrections. Thus, after coarse-graining over oscillations, both the scalar and fermion components redshift approximately as radiation. 

This behaviour is consistent with all oscillatory solutions around the saturation point for odd $p$, in agreement with our arguments in Sec.~\ref{sec:yukawageneral}. However, there is an important practical distinction in matter domination. Since the effective potential decays rapidly, the scalar reaches the saturation point within a reasonable interval of the scale factor only if it starts with a sufficiently large initial velocity or if the source strength $D$ is large enough. For larger odd $p$, the potential is flatter away from the saturation point, in particular near $\phi=0$, and the field may take a long time to reach $\phi_s=-1$ over the range of scale factors considered.

Finally, at the moment of crossing $\phi_s$, the effective fermion mass vanishes, $m_{\rm eff}=0$, and therefore $x=0$. Strictly speaking, the non-relativistic approximation breaks down in a small neighbourhood of each crossing. However, this occurs only over a cosmologically very short time interval. 
Since $|\phi+1|\sim a^{-1}$, the envelope of $x$ remains approximately constant in the saturation regime similar to the relativistic case. Thus, if this constant is large, the non-relativistic approximation remains self-consistent except during the short intervals around the zero crossings. One can check that this is the case by using the exact expressions for $\rho_\psi$ and $p_\psi$ given in Eq.~ \eqref{eq:rho_p_exact_compact}, see also Ref.~\cite{Dom_nech_2021}.

\begin{figure}[t]
\centering
\includegraphics[width=1.05\columnwidth]{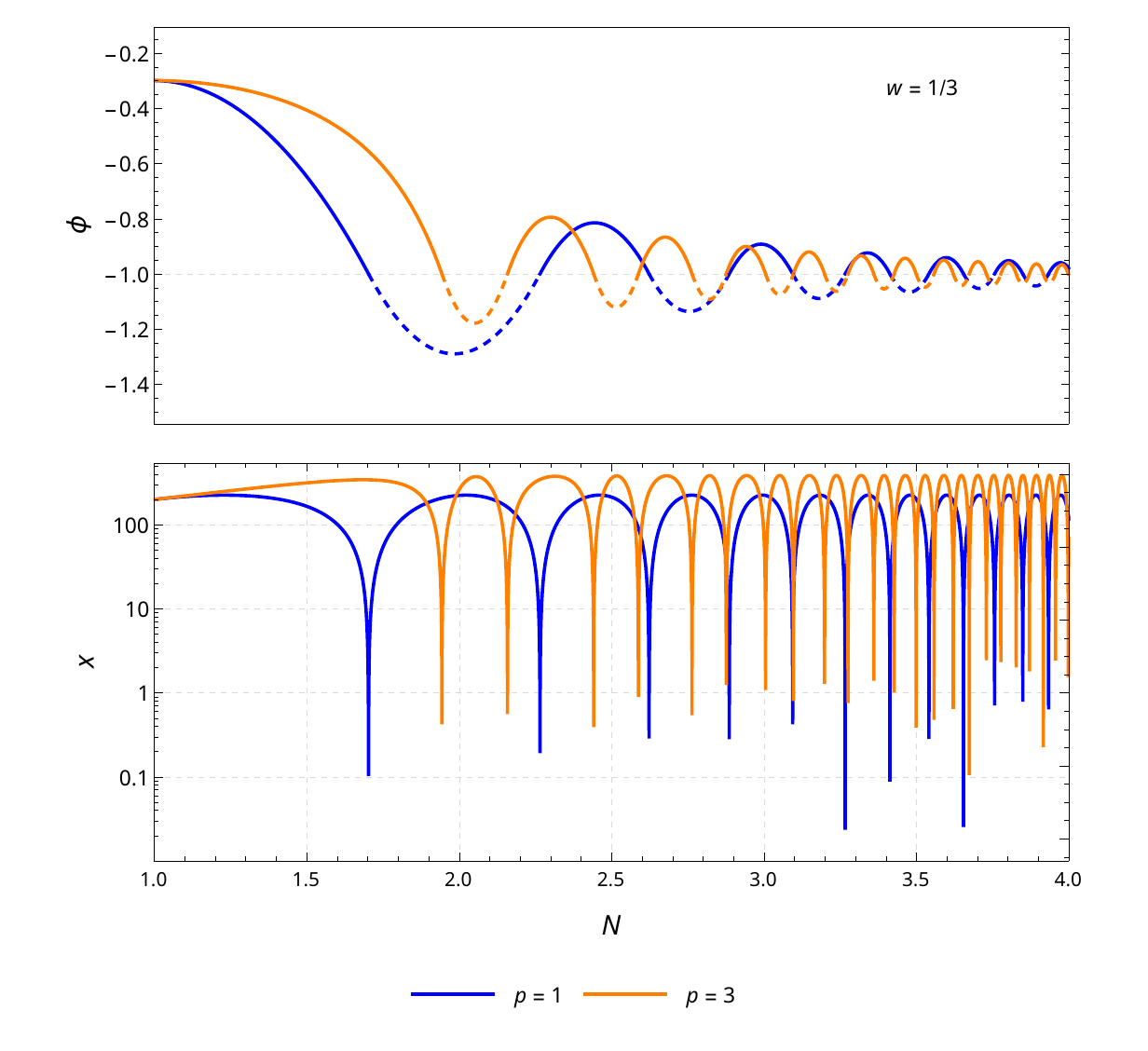}
\caption{
{\bf Upper panel:} Evolution of $\phi$ for $\phi_i=-0.3$ and $\phi'_i=0$ in a radiation-dominated background, $w=1/3$ with $D=1$, for $p=1$ and $p=3$. The change of linestyle indicates the change of sign of the non-relativistic source across the saturation point $\phi_s=-1$.
{\bf Lower panel:} Evolution of the ratio $x$, starting from $x_i=200$. Whenever the field crosses $\phi_s$, one has $x=0$, but the field rapidly moves away from the saturation point and $x$ returns to large values. The envelope of $x$ remains approximately constant, as expected from $x\sim a|1+\phi^p|$ and $|\phi+1|\sim a^{-1}$, see \ref{sec:nonrelplus}.
}
\label{fig:nonrel_p13_w13}
\end{figure}

In Fig.~\ref{fig:nonrel_p13_w13}, we show the evolution of oscillatory solutions for $D=1$ and couplings $p=1$ and $p=3$ in radiation domination, together with the corresponding evolution of $x$. We have tuned $x\sim 10^2$, and on average we find that it remains so, with the visible spikes in the lower panel indicating an extremely brief transition to a relativistic regime before bouncing back to the non-relativistic regime. 

\begin{figure}[t]
\centering
\includegraphics[width=1.05\columnwidth]{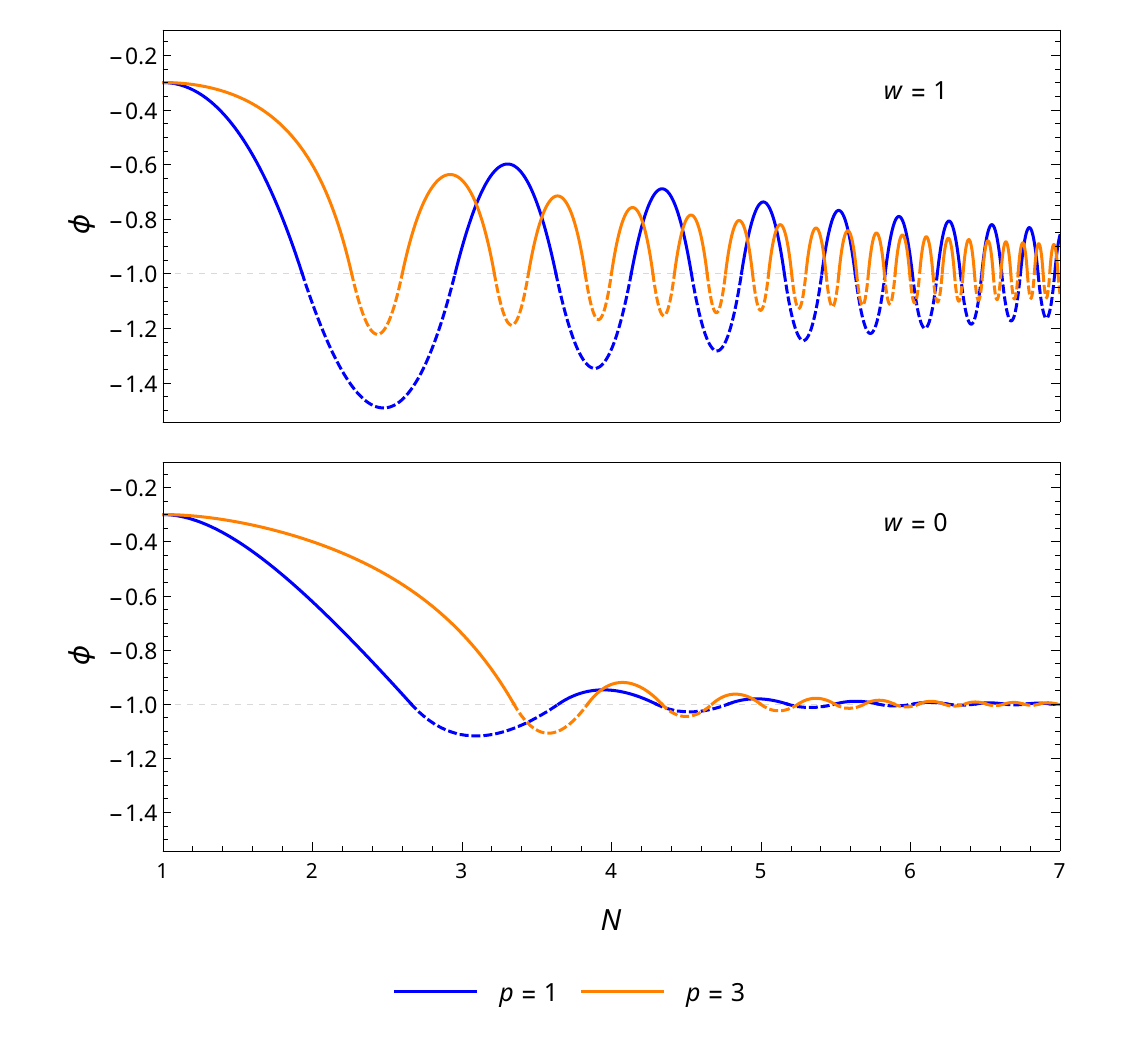}
\caption{{\bf Upper panel:} Evolution of $\phi$ for $\phi_i=-0.3$ and $\phi'_i=0$ in a stiff fluid-dominated background in the non-relativistic limit with $D=1$, for $p=1$ and $p=3$.
{\bf Lower panel:} Same system for matter background. See \ref{sec:nonrelplus}
}
\label{fig:nonrel_p13_w10}
\end{figure}

In Fig. \ref{fig:nonrel_p13_w10} we present the evolution of the same system but for a stiff fluid and matter background. As in the relativistic case, the oscillation amplitude and frequency are larger for a stiff fluid. Additionally, for $w=1$ it is evident that during the transient phase the oscillations around $\phi_s=-1$ are visibly non-symmetric for $p=3$ coupling with the minimum displacement below $\phi_s=-1$ to be smaller than the maximum one above, as the effective potential is also very asymmetric close to the cusp for increasingly higher $p$. However, they quickly converge to the approximate cusp oscillatory solution of Eq. (\ref{eq:nr_oddp_first_harmonic}).


\subsubsection{The even $p$ case\label{subsec:evenpcase}}

For even $p$, the minimum lies at $\phi=0$, and the scalar field equation \eqref{eq:phi_nr_main} becomes
\begin{equation}
    \phi''+\frac{5-3w}{2a}\phi'
    +Dp\,\phi^{p-1}a^{3w-2}=0.
    \label{eq:nr_even_eom}
\end{equation}
Note that, for $f(\varphi)>0$, the even case is qualitatively different from the odd case because of the absence of a saturation point at $m_{\rm eff}=0$. We explore the solutions around $\phi\sim 0$ below.

First, we find that the case $p=2$ is again special. It admits a closed-form solution, with an oscillatory branch for $w=1$ and $w=1/3$, and a decaying asymptotic branch for $w=0$. For $w\neq0$, the oscillatory solution has a closed form in terms of Bessel functions,
\begin{equation}
    \phi(a)=a^{\frac{3(w-1)}{4}}
    \left[c_1J_\nu(z)+c_2Y_\nu(z)\right],\label{eq:solutionphinonrelp2}
\end{equation}
with
\begin{equation}
    \nu=\frac{1-w}{2w},
    \qquad
    z(a)=\frac{2\sqrt{2D}}{3w}a^{\frac{3w}{2}}.
\end{equation}
At late times, that is $a\gg 1 $, Eq.~\eqref{eq:solutionphinonrelp2} asymptotes to
\begin{equation}
    \phi(a)\sim
    A a^{-3/4}
    \cos\left(\frac{2\sqrt{2D}}{3w}a^{\frac{3w}{2}}+\theta\right),
\end{equation}
which differs from the saturation-branch solutions for $p$ odd, see Eq.~\eqref{eq:nr_oddp_first_harmonic}. Here the envelope decays as $a^{-3/4}$, which implies
\begin{equation}
    \rho_\phi\sim a^{-9/2},
    \qquad
    \rho_\psi\sim a^{-3}\,,
\end{equation}
which implies that the fermions redshift more slowly, while the scalar behaviour resembles a stiff fluid with equation of state $w=1/2$. The reason for this difference is that for even $p$ and positive $f(\varphi)$ there is no saturation point and, therefore, the arguments of Sec.~\ref{sec:yukawageneral} do not apply. The solution \eqref{eq:solutionphinonrelp2} oscillates around $\phi\sim 0$ with a different envelope.

For $w=0$, the solution depends on the value of $D$. For $D<9/32$, the solution at late times obeys a power-law in the form of
\begin{equation}
    \phi=c_+a^{r_+}+c_-a^{r_-}\sim c_+a^{r_+},
    \qquad a\gg 1\,
\end{equation}
where
\begin{equation}
    r_\pm=\frac{-3\pm\sqrt{9-32D}}{4}\,.
\end{equation}
For $D\geq9/32$, the late time solution is oscillatory with logarithmic frequency dependence, namely
\begin{equation}
    \phi\sim a^{-3/4}
\cos\!\left(
\frac14\sqrt{32D-9}\,\ln a+\theta
\right)\,.
\end{equation}
For these types of solutions, it is then possible for $w=1/3$ and $w=1$ that, if the fermions start as relativistic, the scalar starts oscillating around $\phi=0$, where the fermions obtain their bare mass. However, the ratio $x$ only grows with the scale factor, and the fermions will gradually turn non-relativistic.

For $p>2$, the late-time behaviour depends strongly on the background. We have identified both oscillatory and asymptotic solutions. 
We find non-harmonic oscillatory solutions which behave at late times as 
    $\phi\sim a^{-\frac{3}{p+2}}\times\text{oscillations}(a)$,
whereas for $w\leq1/3$ one may instead obtain non-oscillatory decays. We find in asymptotic solutions that the field behaves as
\begin{equation}
    \phi\sim a^{-\frac{3w}{p-2}},
    \qquad 0<w\leq\tfrac13,
\end{equation}
and
\begin{equation}
    \phi\sim (\ln a)^{-\frac{1}{p-2}},
    \qquad w=0.
    \label{eq:nr_log_general}
\end{equation}
For a stiff fluid background, the scalar field will always eventually cross the $\phi=0$ point and will end up in the saturation regime. These asymptotic branches are also the same for odd $p$ when the field evolves toward the flat region near $\phi\simeq0$ rather than toward the minimum at $\phi_s=-1$.

\subsection{Relativistic and non-relativistic fermions and $f(\varphi)\propto -\varphi^p$}

Let us now consider the same monomial coupling but with a negative sign for $f(\varphi)$. In this case, the effective mass of the fermions becomes $m_{\rm eff}=|1-\phi^p|$, in terms of the dimensionless field $\phi$. Naively, one expects that this is the same as taking $\phi\rightarrow-\phi$, which is indeed the case for any $p$ odd; however, there is a qualitative difference when $p$ is even. For these reasons, we provide a brief overview of the main differences below without focusing on the details of the equations of motion.

Concretely, in the relativistic regime, the effective potential \eqref{eq:veff}  for even $p$ is now symmetric around $\phi=0$. The origin is a local maximum and the potential forms a double well with minima at $\phi_s=\pm1$, corresponding to \textit{two degenerate saturation points}. This is the most interesting difference for even $p$ as one may generate domain walls for the random initial conditions generated during inflation. We come back to this possibility in Sec.~\ref{sec:Discussion}.

Expanding around either minimum again gives harmonic motion with envelope $a^{-1}$, with asymptotic solution
\begin{equation}
    \phi(a)\approx
    \pm1+\frac{A}{a}
    \cos\left(
    \frac{2p\sqrt{K}}{3w+1}a^{\frac{3w+1}{2}}+\theta_\pm
    \right)\,.
    \label{eq:phi_evenminus_osc}
\end{equation}
In the non-relativistic regime, the effective potential also has the same minima at $\phi=\pm1$. The local dynamics is cusp-like rather than quadratic, but numerically, the late-time envelope again scales as
$|\phi\pm1|_{\rm env}\sim a^{-1}$.

Thus, the sign-flipped even-$p$ branch behaves similarly to the previous odd-$p$ saturation branch. Finally, for odd $p$, the flipped-sign coupling yields an effective potential that is the mirror image of the corresponding $+$ coupling case across $\phi=0$. Therefore, the solutions are qualitatively the same, except that the minimum which minimises the effective mass now lies at $\phi_s=+1$ instead of $\phi_s=-1$. 

\section{Solutions for monomial couplings with $V_{\rm bare}$}
\label{sec:monomial_with_bare}

In this section, we generalise our previous analysis by including an explicit bare self-interaction for the scalar,
to the effective potential \eqref{eq:veff}.

This additional contribution qualitatively modifies the late-time dynamics, because it becomes increasingly important relative to the fermion-induced terms as the universe expands. Depending on the initial conditions, the scalar may either fail to reach the scaling regime or, after reaching it, be driven away from it toward the minimum of the bare potential. We find that this transition is not universal: it depends on the background cosmology, the parameters of the bare potential, and the form of the fermion coupling. It is therefore necessary to understand how the bare potential competes with, and eventually dominates over, the fermion-induced part of the effective potential.

Expressed in terms of the dimensionless field $\phi$, the bare contribution to the effective potential associated with the equation of motion and in Eq.~\eqref{eq:veff}, takes the form
\begin{equation}
    V_{\rm eff}^{\Lambda_q}(\phi\,;\,a)
    =
    \frac{\Lambda_q}{q}a^{3w+1}\phi^q\,,
\end{equation}
with $q>1$.
Although the bare potential eventually dominates the scalar dynamics, the actual late-time minimum depends on competing powers of $\phi$ with respect to $f(\varphi)\propto \pm \varphi^p$ \eqref{eq:fvarphi}. For this reason, we treat the cases $q>p$, $q=p$, and $q<p$ separately.

Overall, the key dynamical feature of including the bare potential is that the late-time minimum of the full effective potential, including the fermion contribution, is no longer fixed at the fermion-induced saturation point we found in the previous sections but instead generally drifts with the scale factor. At early times, provided we require the bare parameter $\Lambda_q\ll K, D$, this minimum can still lie close to the saturation point, while at later times the bare potential pulls the field toward the origin, making the fermions eventually recover their bare mass.

The general picture is therefore similar to the one encountered in the absence of a bare potential, but with an important modification: the field either approaches a \textit{time-dependent minimum} asymptotically or enters an oscillatory regime around it, depending on the parameters and initial conditions. Numerically, we find that, for fixed initial data, increasing the monomial index $p$ tends to delay the onset of oscillations and makes it harder for the field to settle into the saturation regime. Once oscillations begin, however, their frequency grows with $p$, reflecting the increasing steepness of the effective potential around the minimum.

\subsection{Relativistic fermions and $f(\varphi)\propto+\varphi^p$ \label{sec:relfermpoddbare}}

In the relativistic regime, the effective potential \eqref{eq:veff} becomes
\begin{equation}
    V_{\rm eff}^{\rm rel}(\phi\,;\,a)
    =
    K
    \left(\phi^p+\frac12\phi^{2p}\right)a^{3w-1}
    +\frac{\Lambda_q}{q}a^{3w+1}\phi^q .
\end{equation}
The first term is the fermion-induced contribution discussed in the previous section, while the second term is the bare self-interaction. The effective potential has extrema located at
\begin{equation}
    Kp\left(\phi^{p-1}+\phi^{2p-1}\right)
    +\Lambda_q a^2\phi^{q-1}=0\,.
    \label{eq:min_condition_general_compact}
\end{equation}

\begin{figure}[t]
\includegraphics[width=0.5\textwidth]{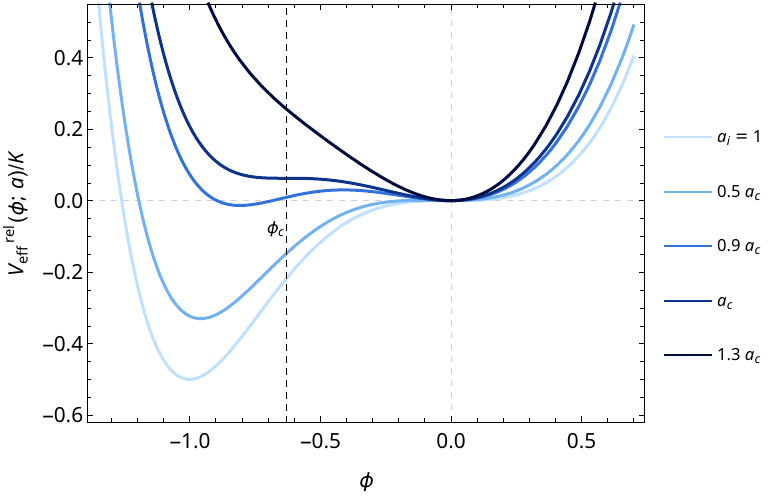}
\caption{The evolution of the relativistic effective potential for $w=1/3$, $p=3$, and $q=2$. The curves show $V_{\rm eff}^{\rm rel}(\phi;a)/K$ at $a_i=1$ and at later values of the scale factor, assuming $\Lambda_q \ll Kp$. Initially, the potential contains a minimum close to the saturation point $\phi_s=-1$, while the quadratic bare term makes $\phi=0$ a shallow local minimum. As the scale factor increases, the off-origin minimum and the intervening local maximum move toward each other and merge at $a=a_c$. For $a>a_c$, the off-origin branch disappears and the late-time potential has a single minimum at $\phi=0$. This illustrates the finite-time transition that occurs for $q<p$; see Sec.~\ref{sec:relfermpoddbare}.
}
\label{fig:pot_p3}
\end{figure}

For even $p$, the only minimum is at $\phi=0$. Therefore, the end state is a solution that approaches this value, as discussed in Sec.~\ref{subsec:evenpcase}. In this case, the presence of the bare potential implies that the late-time oscillations around zero are controlled by the bare scalar mass scale. The fermions are then driven toward the value at which their effective mass equals their bare mass. As the oscillations around $\phi\sim 0$ in the $p$ even case are qualitatively similar to the $p$ odd case, we focus only on the $p$ odd case from now on.

Since we are mainly interested in cases where the scalar first reaches the saturation point and later exits it once the bare potential dominates, we focus on the evolution of the minimum near the saturation point. From Eq.~\eqref{eq:min_condition_general_compact}, we see that, as long as $\Lambda_q a^2\ll Kp$, the bare contribution is negligible, and the nontrivial minimum remains close to the saturation point $\phi_s=-1$.
Expanding Eq.~\eqref{eq:min_condition_general_compact} around $\phi_s$, one finds that at early times
\begin{equation}
    \phi_{\min}(a)\simeq -1+\frac{\Lambda_q}{Kp^2}\,a^2 .
\end{equation}
Thus, the effect of the bare term is to push the minimum slowly away from $-1$ and toward the origin. Let us look at different possibilities for $p$ and $q$ separately.

\subsubsection{The $p=1$ case}

First, for the linear Yukawa coupling case $p=1$, Eq.~\eqref{eq:min_condition_general_compact} reduces to
\begin{equation}\label{eq:minconditionp1}
    K(1+\phi_{\min})
    +\Lambda_q a^2\phi_{\min}^{\,q-1}=0\,.
\end{equation}
At late times, when the minimum has already moved into the regime $|\phi_{\min}|\ll1$, Eq.~\eqref{eq:minconditionp1} gives
\begin{equation}
    |\phi_{\min}(a)|
    \approx
    \left(\frac{K}{\Lambda_q}\right)^{1/(q-1)}
    a^{-2/(q-1)}\,.\label{eq:phi_min_pone}
\end{equation}
For the quadratic bare potential $q=2$, the position of the minimum can be derived exactly, that is
\begin{equation}
    \phi_{\min}(a)= -\frac{K}{K+\Lambda_2 a^2}.
\end{equation}
This expression shows the interpolation between the early-time value $\phi_{\min}\simeq -1$ and the late-time limit $\phi_{\min}\to0$. Importantly, the location of the instantaneous minimum does not depend explicitly on the background cosmological fluid. The background affects instead, how quickly the field reaches and tracks this branch, with the stiff-fluid case typically being the fastest.

\subsubsection{The $p>1$ and $q>p$ case}

For $p>1$ and $q>p$, once the minimum enters the small-field region $|\phi_{\min}|\ll1$, the $\phi^{2p-1}$ term in Eq.~\eqref{eq:min_condition_general_compact} becomes subleading relative to $\phi^{p-1}$. One then finds
\begin{equation}
    |\phi_{\min}(a)|
    \approx
    \left(\frac{Kp}{\Lambda_q}\right)^{1/(q-p)}
    a^{-2/(q-p)}.
    \label{eq:phi_min_qgp}
\end{equation}
Hence, the off-origin minimum survives for all finite $a$, but it drifts continuously toward $\phi=0$, which is approached only asymptotically.

\subsubsection{The $p>1$ and $q<p$ case}

The situation is different when $q<p$. In this case, near the origin, the bare force, which scales as $\phi^{q-1}$, is less suppressed than the fermion-induced force, which scales as $\phi^{p-1}$. For the branch in the interval $-1<\phi<0$, the extremum condition can be written as
\begin{equation}
    \Lambda_q a^2 = -Kp \phi^{p-q}(1+\phi^p)\,.
    \label{eq:extremumconditionpgq}
\end{equation}
For $q<p$, the right-hand side vanishes at both endpoints of the interval, namely at $\phi=-1$ and $\phi=0$, and is positive in between. It therefore has a maximum at an intermediate value of $\phi$. This means that, for sufficiently small $a$, Eq.~\eqref{eq:extremumconditionpgq} has two solutions in the interval $-1<\phi<0$: one corresponds to a local maximum of the effective potential, and the other to a local minimum close to the saturation point. This behaviour is illustrated in Fig. \ref{fig:pot_p3} for the representative case $p=3$, $q=2$ in radiation domination.

As the scale factor increases, the left-hand side of
Eq.~\eqref{eq:extremumconditionpgq} grows. The two extrema, therefore, move toward each other. The critical point is reached when they merge. To find this point, define
\begin{equation}
    R(\phi) = - Kp \phi^{p-q}(1+\phi^p),
    \qquad
    -1<\phi<0\,.
\end{equation}
The critical field value $\phi_c$ is determined by $ R'(\phi_c)=0$.
One then obtains that,
\begin{equation}
    |\phi_c|^p = \frac{p-q}{2p-q}\,.
\end{equation}
Substituting this value back into Eq.~\eqref{eq:extremumconditionpgq}
gives the critical scale factor
\begin{equation}
    a_c^2
    =
    \frac{Kp^2}{\Lambda_q(p-q)}
    \left(
        \frac{p-q}{2p-q}
    \right)^{(2p-q)/p}.
    \label{eq:ac_rel_q_less_p}
\end{equation}

Thus, for $a<a_c$, the effective potential contains an off-origin minimum near $\phi=-1$, separated from the origin by a local maximum. At $a=a_c$, this minimum and the intervening maximum merge and annihilate. For $a>a_c$, no nonzero minimum remains, and the only minimum is at the origin. The scalar is therefore driven into origin-centered oscillations.

Lastly, for $p=q$ as $q$ is defined strictly even and therefore for $p$ even, we have that $\phi_\text{min}=0$ always, so what can happen is that the field will transit to oscillations with mass scale the bare potential mass.

\begin{figure}[t]
    \centering
    \includegraphics[width=0.5\textwidth]{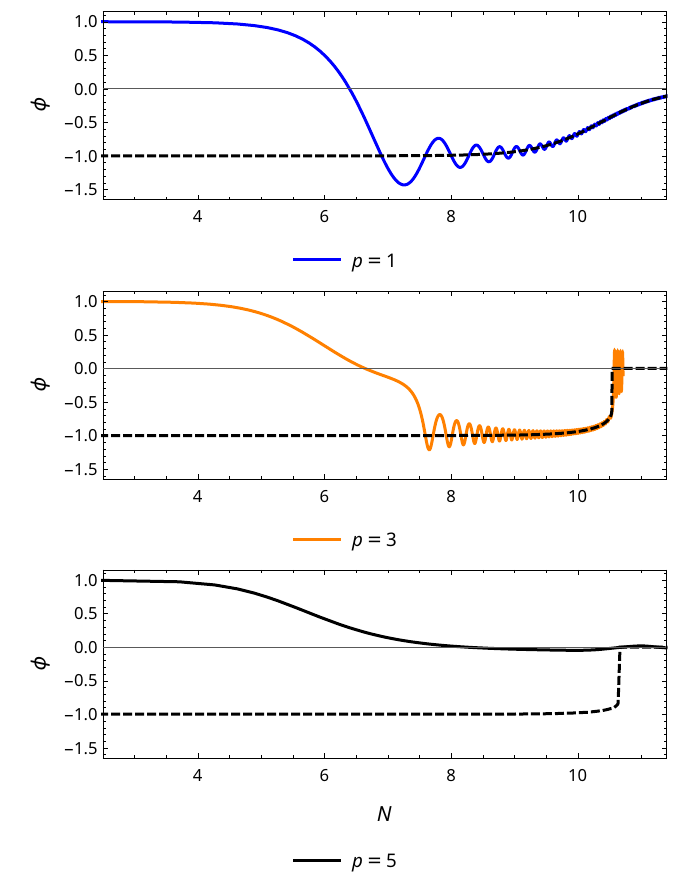}
    \caption{Evolution for $\phi_i=+1$ and $\phi'_i=0$, with $w=1/3$, $q=2$, and $p=1$ (\textbf{top panel}), $p=3$ (\textbf{middle panel}), and $p=5$ (\textbf{lower panel}) in the relativistic regime. The solution first enters the saturation regime near $\phi \approx -1$ and then oscillates around the time-dependent minimum of the effective potential. Initially, the displacement of this minimum from the saturation point grows as $\delta\phi_{\min}\sim a^2$, and the minimum eventually moves toward $\phi=0$. The dashed line denotes the evolution of the time-dependent minimum. The solution tracks this minimum for $p=1$ and $p=3$, but not for $p=5$, where the scalar field does not reach the saturation regime and instead starts oscillating around $\phi=0$, see Sec. \ref{sec:relfermpoddbare}.}
    \label{fig:pot_p1}
\end{figure}

\subsubsection{Numerical exploration}

These results make clear that the bare potential does not simply add a small correction to the relativistic dynamics. Rather, it changes the asymptotic structure of the problem. In particular, the fermion-induced minimum near $\phi=-1$ is only transient. At late times, the field is either forced to track a minimum that drifts toward the origin, or the off-origin branch disappears at a finite scale factor, and the field subsequently evolves around $\phi=0$. We present examples in Fig.~\ref{fig:pot_p1}, using the same initial conditions as in the previous section and assuming radiation domination. 

\begin{figure}[t]
    \centering
    \includegraphics[width=0.5\textwidth]{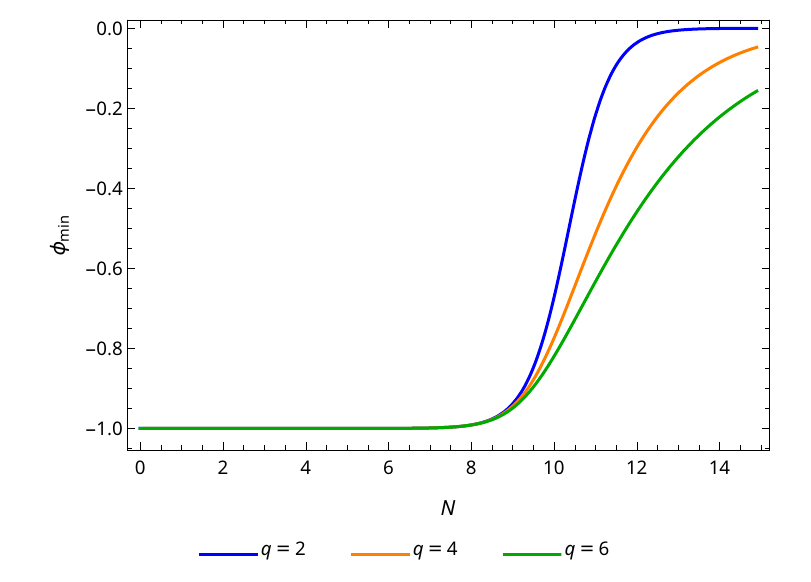}
    \caption{The drift of the effective minimum for different values of $q=2$ (blue line), $q=4$ (orange line) and $q=6$ (green line) for $p=1$ linear coupling in the relativistic regime. The effective minimum approaches $\phi=0$ at late times as $\phi(a)_\text{min}\sim a^{-2/(q-p)}$, which implies that for fixed $K=10^{-5}$, $\Lambda_q = 10^{-14}$ parameters, the effective minimum drifts the fastest for the quadratic case $q=2$. The scalar field oscillates around this drifting minimum, see Sec. \ref{sec:relfermpoddbare}.}
    \label{fig:mindrift}
\end{figure}

For the linear Yukawa case, $p=1$, the field reaches the saturation regime, oscillates around the effective minimum, and then tracks this minimum as it approaches zero asymptotically. For $p=3$ and $q=2$, the field initially approaches the off-origin minimum, but this branch terminates at finite $a=a_c$ when it merges with the intervening maximum. After this transition, the only remaining minimum is at $\phi=0$. Finally, for $p=5$, the field does not reach the saturation point for the chosen initial conditions and instead begins oscillating around $\phi=0$. In this case, because $p>q$, the origin is already a competing minimum before the off-origin branch disappears. For fixed $p$ and varying $q$, the qualitative behaviour is similar. The main difference is that the drift of the effective minimum toward the origin becomes slower for larger $q$, as shown in Fig.~\ref{fig:mindrift}.

\subsection{Relativistic fermions and $f(\varphi)=-\varphi^p$ ($p$ even) \label{sec:relativisticfermionbaresection2}}

For the sign-flipped even-$p$ branch, the relativistic effective potential \eqref{eq:veff} becomes
\begin{equation}\label{eq:veffrelbare}
    V_{\rm eff}^{\rm rel}(\phi\,;\,a)
    =
    K
    \left(-\phi^p+\frac12\phi^{2p}\right)a^{3w-1}
    +\frac{\Lambda_q}{q}a^{3w+1}\phi^q .
\end{equation}
In the absence of the bare term, as we saw in the previous section, this potential possesses two symmetric minima near $\phi=\pm1$. The role of the bare contribution is again to pull both minima inward, eventually making the origin the preferred vacuum.

The off-origin extrema of the effective potential \eqref{eq:veffrelbare} are given by
\begin{equation}
    Kp\left(\phi^p-1\right)
    +\Lambda_q a^2\phi^{q-p}=0.
    \label{eq:nonzero_rel_minus_even_compact}
\end{equation}
At early times, when the bare term is still subdominant, the two minima are displaced only slightly from their induced values, namely
\begin{equation}
    \phi_{\min}^{(\pm)}(a)
    \simeq
    \pm 1\mp\frac{\Lambda_q}{Kp^2}a^2\,,
\end{equation}
when $\phi\sim 1$.
Thus, both minima move symmetrically toward the origin as the universe expands. Unless $\phi$ starts exactly at the origin, the scalar field eventually falls into one of the two minima and oscillates around the corresponding drifting minimum. 

As we show below, the distinction between $q>p$, $q=p$, and $q<p$ is physically important. For $q>p$, the late-time dynamics are governed by two minima that survive and slowly approach the origin. For $q\le p$, by contrast, the off-origin structure terminates at a finite scale factor, so that the scalar necessarily undergoes a transition to evolution around $\phi=0$. We study the evolution of these effective minima in more detail for the cases $q>p$, $q=p$, and $q<p$ separately.

\subsubsection{The $q>p$ case}

If $q>p$, the two distinct minima are going to collapse to a single minimum asymptotically. At late times, they move into the small-field region around $\phi=0$. Expanding Eq.~\eqref{eq:nonzero_rel_minus_even_compact} for $\phi\sim 0$ we find that the minima are located at
\begin{equation}
    |\phi_{\min}(a)|
    \approx
    \left(\frac{Kp}{\Lambda_q}\right)^{1/(q-p)}
    a^{-2/(q-p)}.
    \label{eq:phi_min_rqgp}
\end{equation}
They therefore approach the origin asymptotically, as in the odd-$p$ case of Sec.~\ref{sec:relfermpoddbare}.

\subsubsection{The $q=p$ case}

\begin{figure}[t]
    \centering
    \includegraphics[width=0.5\textwidth]{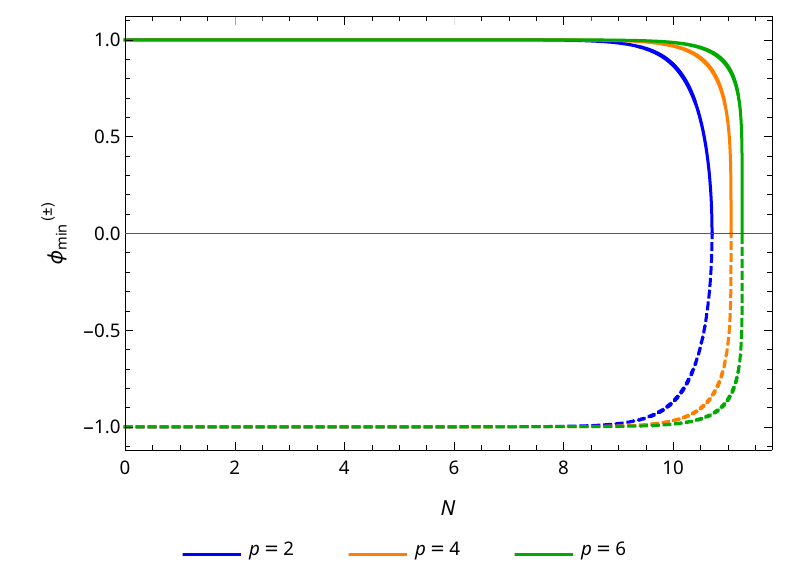}
    \caption{The drift of the effective minima for different values of $p=q=2$ (blue lines), $p=q=4$ (red lines) and $p=q=6$ (green lines) in the relativistic regime. The solid lines denote the positive minimum, and the dashed lines the negative one. The minima merge at later times for higher powers. The scalar field evolves oscillating around those drifting minima, see Sec. \ref{sec:relativisticfermionbaresection2}.}
    \label{fig:mindriftpq}
\end{figure}

When $q=p$, the behaviour is qualitatively different because the scale-factor dependence in Eq.~\eqref{eq:nonzero_rel_minus_even_compact} enters only through the overall coefficient. In this case, one finds the following exact result for the position of the effective minima, 
\begin{equation}
    \phi_{\min}^{(\pm)}(a)
    =
   \pm\left( 1-\frac{\Lambda_q}{Kp}a^2\right)^{1/p}\qquad a\le a_c\,,
\end{equation}
which shows that the two minima move inward and merge at the origin at a critical scale factor given by
\begin{equation}
    a_c^2=\frac{Kp}{\Lambda_q}\,.
\end{equation}
Thus, the transition occurs at a finite scale factor and is controlled directly by the ratio of the induced to the bare terms. As shown in Fig. \ref{fig:mindriftpq}, the factor $p$ makes the convergence of the minima slower with increasing values of $p$ and $q$.

In passing, we note that the case $p=q$ is interesting from the perspective of phase transitions. Assuming that the scalar fields start at $\phi=0$, which is an unstable point and exhibits small fluctuations, then some regions might fall into the positive and others into the negative minimum, but as we have shown, the two degenerate minima converge and eventually disappear.

\subsubsection{The $q<p$ case}

For $q<p$, the off-origin minima disappear even earlier. In this case, the two minima merge with the neighbouring maxima and are annihilated at a critical point, in the same manner as in the previous odd-$p$ branch. Beyond this point, the origin is the only remaining minimum of the potential.

\subsection{Non-relativistic fermions}

The dynamics of the scalar field for non-relativistic fermions have the same broad physical picture as in the relativistic regime: the bare potential eventually dominates and drives the system toward the origin. The main difference is that, because of the absolute value of the effective mass, the effective potential has a cusp at the saturation point. We find that the transition from saturation to origin-centred evolution occurs through the disappearance of cusp minima at finite $a_c$. Below, we focus on the $p$ odd case for $f(\varphi) = \varphi^p$, since the case when $f(\varphi) =-\varphi^p$ amounts to a flip of the sign of $\varphi$. We later focus on the $p$ even case for $f(\varphi) =-\varphi^p$, which is the case that contains a saturation point and a richer structure of the effective potential.

\subsubsection{The $p$ odd case}\label{subsubec:poddnonrel}

In the non-relativistic regime, for $f(\varphi)\propto+\varphi^p$, the induced contribution is no longer smooth, and the effective potential \eqref{eq:veff} becomes
\begin{equation}\label{eq:veffnonrelbareodd}
    V_{\rm eff}^{\text{non-rel}}(\phi\,;\,a)
    =
    D\,a^{3w-2}|1+\phi^p|
    +\frac{\Lambda_q}{q}a^{3w+1}\phi^q .
\end{equation}
Note that the absolute value produces a cusp in the effective potential at the saturation point $\phi=-1$, so the minimum inherited from the fermion sector is initially cusp-like rather than analytic. This is the main qualitative difference from the relativistic regime of Sec.~\ref{sec:relfermpoddbare} and \ref{sec:relativisticfermionbaresection2}.

The aforementioned cusp at $\phi=-1$ remains a local minimum only as long as the bare-potential slope does not overcome the fermion-induced one-sided slope. This is the case whenever $a^3<\frac{Dp}{\Lambda_q}$. If so, the transition to the bare potential dominated regime occurs at
\begin{equation}
    a_c^3=\frac{Dp}{\Lambda_q}.
    \label{eq:ac_nr_plusodd_compact}
\end{equation}

For $a<a_c$, the dynamics is still dominated by the cusp and the scalar is efficiently attracted toward the saturation point. Once $a$ exceeds $a_c$, however, the cusp minimum disappears, and one must instead look for smooth extrema inside the interval $-1<\phi<0$. In this region, the effective potential \eqref{eq:veffnonrelbareodd} extrema lie at
\begin{equation}
    Dp\,\phi^{p-1}+\Lambda_q a^3\phi^{q-1}=0.
\end{equation}
For $q>p$, after $a>a_c$ the minimum starts drifting towards zero
\begin{equation}
    |\phi_{\min}(a)| = \left(\frac{a}{a_c}\right)^{-3/(q-p)}\,.
    \label{eq:phi_min_nr_plusodd_compact}
\end{equation}
, Eq.~\eqref{eq:phi_min_nr_plusodd_compact}, shows that the minimum survives beyond the cusp transition and moves gradually toward the origin, in close analogy with the relativistic tracking branches of Sec.~\ref{sec:relfermpoddbare} and \ref{sec:relativisticfermionbaresection2}, although with a different scaling in $a$.

For $q<p$, the situation changes qualitatively. After the cusp disappears, there is no longer any smooth off-origin extrema that can be continuously connected to the earlier minimum. The branch therefore terminates at finite $a_c$, after which $\phi=0$ becomes the unique minimum. In other words, at some point the scalar field exits the saturation point and directly oscillates around $\phi\sim 0$.

\subsubsection{The $p$ even case}

For the sign-flipped even-$p$ branch, that is, for $f(\varphi)\propto -\varphi^p$ (which is the one allowing for a saturation point), the non-relativistic effective potential \eqref{eq:veff} reads
\begin{equation}
    V_{\rm eff}^{\text{non-rel}}(\phi\,;\,a)
    =
    D\,a^{3w-2}|1-\phi^p|
    +\frac{\Lambda_q}{q}a^{3w+1}\phi^q\,.
\end{equation}
In this case, the fermion-induced term generates cusp minima at $\phi=\pm1$. The drift of the effective minima is identical to the $f(\varphi)\propto+\varphi^p$ case for $p$ odd, in \ref{subsubec:poddnonrel} with the two minima merging or approaching depending on $p$ and $q$ in exactly the same way after the $a_c$.

Specifically, for $q>p$ we have the same behaviour again as in the odd case, where after $a>a_c$ the minimum drifts towards $\phi=0$ in the same way as in the odd case.
For $q=p$, the competition is instead encoded in the coefficient of $\phi^p$ inside $|\phi|<1$. At $a=a_c$, this coefficient changes sign, so the origin becomes the unique minimum for $a>a_c$. Lastly, if $q<p$, no smooth off-origin minima remain once the cusp minima have disappeared, and again the only late-time minimum is at $\phi=0$. Finally, for $p>q$ as in the $f(\varphi)\propto+\varphi^p$ case, for $a>a_c$, the two minima disappear, and the only minimum is at $\phi=0$.

\subsection{Energy-density scalings}
\label{subsec:energy_densities_bare}

\begin{figure}[t]
    \centering
    \includegraphics[width=0.5\textwidth]{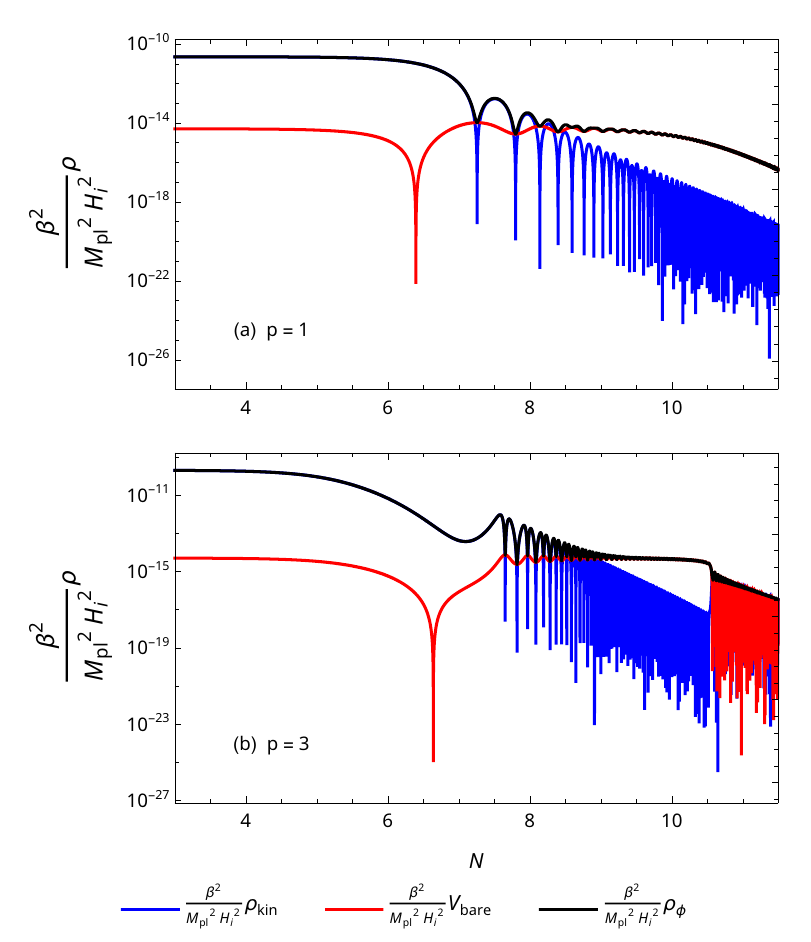}
    \caption{Evolution of the scalar energy-density contributions in the relativistic regime for a quadratic bare potential, $q=2$, with $K=10^{-5}$, $\Lambda_q=10^{-14}$, $\phi_i=1$ and $\phi_i'=0$. The upper and lower panels correspond to $p=1$ and $p=3$, respectively. The blue, red and black curves show the kinetic contribution, the bare-potential contribution and the total scalar energy density. The sharp dips in the kinetic and potential components correspond to turning points and zero-crossings of the oscillatory motion and are accentuated on the logarithmic scale; see Sec.\ref{subsec:energy_densities_bare}.}
    \label{fig:pot_rho}
\end{figure}

So far, we have focused on the scalar field dynamics. However, the behaviour of its energy density \eqref{eq:rhovarphidefinition} is important for the subsequent cosmological expansion. Here, we investigate the scale dependence of the scalar energy density in the presence of a bare potential. 

Note that the energy density $\rho_\varphi$ \eqref{eq:rhovarphidefinition} receives in general two distinct contributions. The first is the energy associated with the slow drift of the effective minimum, while the second is the energy stored in oscillations around this instantaneous minimum. When the effective minimum evolves adiabatically, these two motions can be separated as
\begin{equation}
    \phi(a)
    =
    \phi_{\min}(a)
    +
    {\phi}_{\rm osc}(a)\,,
    \label{eq:adiabatic_split_phi}
\end{equation}
where $\phi_{\min}$ describes the slow motion of the minimum and ${\phi}_{\rm osc}$ is the oscillatory displacement around it. In the relativistic limit, we expect a similar behaviour to the oscillatory solutions in the limit where $V_\text{bare}=0$.

We find that the kinetic energy associated purely with the drift of the minimum is parametrically small. For instance, in previous subsections we found that $|\phi_{\min}(a)|\sim a^{-\gamma}$, where $\gamma$ is a function of $p$ and $q$ (see, e.g., Eqs.~\eqref{eq:phi_min_qgp}, \eqref{eq:phi_min_pone}, 
\eqref{eq:phi_min_rqgp}, \eqref{eq:phi_min_nr_plusodd_compact}). This means that $\phi_{\min}'(a)\sim a^{-\gamma-1}$ and so it follows that
\begin{equation}
    \rho_{\rm kin}^{(\min)}
    \sim
    a^{2-3(1+w)}\left(\phi_{\min}'\right)^2
    \sim
    a^{-3(1+w)-2\gamma}.
\label{eq:rho_kin_min_general_phi}
\end{equation}
By contrast, the bare potential energy evaluated on the drifting minimum scales as
\begin{equation}
    V_{\rm bare}(\phi_{\min})
    \sim
    \phi_{\min}^q
    \sim
    a^{-q\gamma}.
    \label{eq:rho_pot_min_general_phi}
\end{equation}
Since we assumed $q>2$, the kinetic energy induced by its drift is suppressed whenever the minimum moves only on a Hubble time scale.  This shows that the late-time behaviour of $\rho_\varphi$ should not be interpreted simply from $\rho_{\rm kin}^{(\min)}$, but rather from the combination of the slow drift and the oscillatory displacement where the latter dominates the kinetic contribution. Eventually, the kinetic energy decays away, and the bare potential dominates.

We illustrate the aforementioned behaviour in Fig.~\ref{fig:pot_rho} for  relativistic fermions with a quadratic bare potential, $q=2$, using the same initial parameters $K=10^{-5}$, $\Lambda_q=10^{-14}$, $\phi_i=1$ and $\phi_i'=0$ in radiation domination as previous figures. At early times, the scalar energy density is dominated by the kinetic contribution, generated as the field moves towards the effective minimum from the Yukawa coupling. Eventually, the bare-potential contribution becomes important and controls the subsequent evolution. The total relevant quantity is the smooth envelope of the total scalar energy density, shown in black.

From the upper panel of Fig.~\ref{fig:pot_rho}, we see that the effective minimum approaches the origin only asymptotically for $p=1$. In the relativistic quadratic case, we find that the drift of the minimum behaves as 
$\phi_\text{min}\sim a^{-2}$. Therefore, at late times
\begin{equation}
    V_{\rm bare}(\phi_{\min})
    \sim
    \phi_{\min}^2
    \sim
    a^{-4}.
    \label{eq:p1_q2_vbare_scaling}
\end{equation}
Interestingly, this is the same scaling obtained in the relativistic system without a bare potential. In this case, the field continues to follow the slowly drifting minimum, with small oscillations superimposed on top of it.

The situation is different for $p=3$, see bottom panel of Fig.~\ref{fig:pot_rho}. In this case, the off-origin minimum reaches the origin at a finite time. Once this happens, the scalar no longer tracks a displaced minimum. Instead, it behaves as a standard oscillator around the bare quadratic minimum at $\phi=0$, $\langle\rho_\phi\rangle\sim a^{-3}$. Thus, the $p=3$ branch crosses over to a matter-like decay of a scalar condensate in a quadratic potential. 

The contrast between the two panels of Fig.~\ref{fig:pot_rho} therefore reflects the different fate of the effective minimum: for $p=1$ it approaches zero only asymptotically and preserves the radiation-like scaling, while for $p=3$ it disappears into the origin and the subsequent evolution is governed by ordinary quadratic oscillations which on average behave like matter.

Lastly, notice that for the case of $p=1$ with quadratic bare potential, when the fermions eventually become non-relativistic $\phi_\text{min}\sim a^{-3}$ and therefore $V_\text{bare}\sim a^{-6}$, which implies that the oscillations around this effective minimum become relevant at some point.

\section{Summary of the results}
\label{sec:Summary}

In Secs.~\ref{sec:monomial_results} and \ref{sec:monomial_with_bare}, we presented an exhaustive analysis of various cases. Since the analysis may be dense at times, we present here a compact summary of the main results for the busy reader.

We investigated the dynamics of long-range Yukawa forces in the early Universe, arising from Yukawa-type couplings between scalar and fermion fields in a dark sector. We assumed that the fermions can be described as a Fermi gas and that the scalar-fermion system is decoupled from the dominant background cosmic fluid. Within this setup, we studied the behaviour of the solutions for different couplings, background equations of state, and bare scalar potentials. Our main results are:
\begin{enumerate}
    \item For both Yukawa and dilatonic couplings admitting a saturation value $\phi_s$ at which $m_{\rm eff}\rightarrow0$, we find scaling solutions around the saturation point with $|\phi(a)-\phi_s|\propto a^{-1}$ in both the relativistic and non-relativistic limits, independently of the background cosmological fluid. In all cases considered, the leading energy-density scaling is $\rho_\phi\sim\rho_\psi\sim a^{-4}$.

    \item For higher-order couplings, there can also exist asymptotic solutions whose late-time decay depends both on the form of the coupling and on the background cosmological fluid.

    \item In the saturation regime, relativistic fermions remain relativistic, while non-relativistic fermions remain non-relativistic on average, apart from the brief intervals near the zero-mass crossings. In this regime, the coarse-grained ratio satisfies $\langle\rho_\phi\rangle/\langle\rho_\psi\rangle\sim {\rm constant}$.

    \item A bare scalar potential eventually dominates the scalar dynamics and drives the field towards the minimum of the bare potential. As a result, the fermions recover their bare mass and eventually become non-relativistic. The scalar then oscillates around the drifting minimum of the effective potential. The details of this drift depend on the coupling and on the parameters of the bare potential, but are not sensitive to the background cosmological fluid.
\end{enumerate}
In Table~\ref{tab:summary_main_results}, we present a more detailed summary for the interested reader, including the solutions for the different couplings and values of $w$.

\section{Outlook and Discussion}\label{sec:Discussion}

Our work generalised the analysis of Ref.~\cite{Dom_nech_2021} to a general form of the Yukawa-like coupling to fermions in arbitrary cosmological background expansions. As we mainly focus on the background analytical solutions, there are several interesting extensions.

First, we mainly studied the behaviour of the scalar-fermion system near the saturation point and its asymptotics. However, it would be interesting to study the system up to today, in particular, considering the possibility that the fermions constitute the dark matter. We have not pursued this direction in this work because the transition from a general cosmological expansion to the standard radiation-dominated era introduces additional parameters into the model. We plan to explore interesting consequences in the future.

Second, for simplicity, we restricted ourselves to the case of a degenerate Fermi gas. This provides a simple and analytically tractable description, but it would also be interesting, from a cosmological point of view, to study the finite-temperature case. Nevertheless, based on the analysis of Ref.~\cite{Dom_nech_2021}, we expect the thermal case to be qualitatively similar.

Third, we focused on the homogeneous scalar and fermion solutions with fixed fermion number. However, it has been shown that subhorizon perturbations in such a system can lead to an exponential growth of the density contrast \cite{Flores:2020drq,Dom_nech_2023}, to dark-matter halo formation~\cite{Dom_nech_2023}, and potentially PBHs \cite{Flores:2020drq}. It would be interesting to extend this study to perturbations in general cosmological backgrounds, in particular to the study of fermion perturbations near the saturation points and gravitational wave generation \cite{Flores:2022uzt}.

It would also be interesting to investigate whether similar mechanisms could lead to the formation of fermionic halos or other compact fermion objects in the present setup \cite{Del_Grosso_2023,Lu_2025}. Going beyond perturbation theory, such scenarios could also be studied with numerical relativity simulations, which would allow one to capture the strong-gravity dynamics and possible gravitational collapse to black holes~\cite{Aurrekoetxea_2025}.

Lastly, we have found that even couplings can yield multiple degenerate saturation points that appear as degenerate minima in the scalar effective potential. One such realisation is for negative Yukawa $f(\varphi)\propto -\varphi^2$ and quadratic bare potential namely $p=q=2$ as seen in Sec. \ref{sec:relativisticfermionbaresection2}. If the scalar field contains fluctuations, similar to what occurs during inflation, different regions of space could settle into different minima. Such a scalar-fermion system may therefore give rise to domain walls in the early Universe. In the presence of a bare scalar-field potential, the degenerate minima eventually merge, and the domain walls would melt away, as the fermions dilute due to the expansion. The possibility of generating domain walls is interesting, as they may produce a substantial gravitational-wave background (see, e.g., Refs.~\cite {Saikawa_2017,Babichev_2023,dankovsky2025numericalanalysismeltingdomain}), providing an additional signature of our model. We leave a detailed study for future work.

\acknowledgments

We thank Chunshan Lin, Nathaniel Sherrill, Malcolm Fairbairn and Eugene A. Lim for useful discussions. This work is partly supported by the DFG under the Emmy-Noether program, project number 496592360. GD is also supported by the JSPS KAKENHI grant No. JP24K00624. PG is also supported by a Research Project Grant RPG-2021-423 from Leverhulme Trust.

\begin{table*}[t!]
\centering
\scriptsize
\renewcommand{\arraystretch}{1.05}
\setlength{\tabcolsep}{3pt}
\begin{tabular}{p{0.15\textwidth} p{0.23\textwidth} p{0.30\textwidth} p{0.25\textwidth}}
\hline\hline
\textbf{Coupling/regime} &
\textbf{Attractor branch} &
\textbf{Late-time behaviour of the scalar} &
\textbf{Energy-density behaviour}
\\
\hline

\multicolumn{4}{c}{\textit{Monomial coupling \(f(\phi)\propto +\phi^p\), with negligible bare potential}} \\
\hline

Relativistic, odd \(p\) &
Saturation branch at \(\phi_s=-1\), or asymptotic branch near \(\phi=0\) for sufficiently flat potentials &
On the saturation branch,
\(\displaystyle
\phi+1\simeq
\frac{A_p}{a}
\cos\!\left[
\frac{2p\sqrt K}{3w+1}
a^{\frac{3w+1}{2}}+\theta_p
\right].
\)
For \(p>2\), the field may instead enter the asymptotic branch
\(\displaystyle
\phi\sim a^{-\frac{3w+1}{p-2}}
\)
for \(w\leq 1/3\). &
On the saturation branch,
\(\displaystyle
\rho_\phi\sim a^{-4},\quad
\rho_\psi\sim a^{-4},\quad
\rho_\phi/\rho_\psi\sim {\rm const.}
\)
On the asymptotic branch the field approaches \(\phi=0\), so \(m_{\rm eff}\to m_\psi\).
\\[7mm]

Relativistic, even \(p\) &
Minimum at \(\phi=0\), or asymptotic approach to \(\phi=0\) &
For \(p=2\),
\(\displaystyle
\phi\sim
\frac{A}{a}
\cos\!\left[
\frac{2\sqrt{2K}}{3w+1}
a^{\frac{3w+1}{2}}+\theta
\right].
\)
For \(p>2\), asymptotic branches satisfy
\(\displaystyle
\phi\sim a^{-\frac{3w+1}{p-2}}
\)
for \(w\leq 1/3\). &
On oscillatory branches the scalar redshifts approximately as radiation while the fermions remain relativistic. On asymptotic branches the field approaches \(\phi=0\), so \(m_{\rm eff}\to m_\psi\).
\\[7mm]

Non-relativistic, odd \(p\) &
Cusp-like saturation branch at \(\phi_s=-1\) &
\(\displaystyle
|\phi+1|\sim a^{-1}\times \text{parabolic osc.}
\)
&
\(\displaystyle
\rho_\psi\sim a^{-3}|1+\phi^p|\sim a^{-4},
\qquad
\rho_\phi\sim a^{-4}\,.
\)
Thus the coupled sector redshifts as radiation even though the fermions are non-relativistic.
\\[6mm]

Non-relativistic, \(p=2\) &
Minimum at \(\phi=0\) &
\(\displaystyle
\phi\sim
A a^{-3/4}
\cos\!\left[
\frac{2\sqrt{2D}}{3w}
a^{\frac{3w}{2}}+\theta
\right]\,,\text{for}\,\, w>0.
\)
For \(w=0\),
\(\displaystyle
\phi\sim c_+ a^{r_+},
\quad
r_+=\frac{-3+\sqrt{9-32D}}{4}
\)
when \(D<9/32\), while
\(\displaystyle
\phi\sim a^{-3/4}
\cos\!\left[
\frac14\sqrt{32D-9}\,\ln a+\theta
\right]
\)
when \(D>9/32\). &
For \(w>0\), and for the logarithmically oscillatory \(w=0\) branch,
\(\displaystyle
\rho_\psi\sim a^{-3},
\quad
\rho_\phi\sim a^{-9/2},
\quad
\rho_\phi/\rho_\psi\sim a^{-3/2}.
\)
For \(w=0\) and \(D<9/32\),
\(\displaystyle
\rho_\phi/\rho_\psi\sim a^{2r_+}.
\)
\\[9mm]

Non-relativistic, even \(p>2\) &
Asymptotic branch toward \(\phi=0\), except in stiff backgrounds &
\(\displaystyle
\phi\sim
a^{-\frac{3}{p+2}}
\times\text{oscillations}(a)
\quad (w=1),
\)
\(\displaystyle
\phi\sim a^{-\frac{1}{p-2}}
\quad (w=1/3),
\)
and
\(\displaystyle
\phi\sim
(\ln a)^{-\frac{1}{p-2}}
\quad (w=0).
\)
&
\(\displaystyle
\rho_\psi\sim a^{-3},
\qquad
m_{\rm eff}\to m_\psi .
\)
For \(w=1/3\),
\(\displaystyle
\rho_\phi/\rho_\psi
\sim a^{-1-\frac{2}{p-2}},
\)
while for \(w=0\),
\(\displaystyle
\rho_\phi/\rho_\psi
\sim
(\ln a)^{-\frac{2(p-1)}{p-2}}.
\)
\\[8mm]

\hline
\multicolumn{4}{c}{\textit{Monomial coupling \(f(\phi)\propto -\phi^p\)}} \\
\hline

Odd \(p\) &
Mirror saturation branch at \(\phi_s=+1\) &
The dynamics follows from the \(+\phi^p\), odd-\(p\) saturation branch by \(\phi\to-\phi\). &
Same scaling as the odd-\(p\) saturation branch:
\(\displaystyle
\rho_\phi\sim a^{-4},
\qquad
\rho_\psi\sim a^{-4}.
\)
\\[4mm]

Even \(p\) &
Two degenerate saturation points at \(\phi_s=\pm1\) &
Near either minimum,
\(\displaystyle
|\phi\mp1|_{\rm env}\sim a^{-1}
\)
in the saturation regime. &
Both branches correspond to \(m_{\rm eff}\to0\). Hence
\(\displaystyle
\rho_\phi\sim a^{-4},
\qquad
\rho_\psi\sim a^{-4}.
\)
\\[5mm]

\hline
\multicolumn{4}{c}{\textit{Effect of a bare scalar potential}} \\
\hline

$V_{\rm bare}=\lambda_q\phi^q/q$ &
Competition between the fermion-induced branch and the origin &
The bare contribution in the scale-factor effective potential grows as    
\(\displaystyle
V_{\rm eff}^{\Lambda_q}
=
\frac{\Lambda_q}{q}a^{3w+1}\phi^q .
\)
It eventually pulls the scalar toward \(\phi=0\). &
The saturation/scaling regime can be transient. At late times \(m_{\rm eff}\to m_\psi\), and the fermions redshift as non-relativistic matter:
\(\displaystyle
\rho_\psi\sim a^{-3}.
\)
\\[6mm]

\hline\hline
\end{tabular}
\caption{
Summary of the homogeneous late-time solutions of the scalar-fermion system for monomial couplings. 
The dynamics is organised by the sign of the coupling, the parity of $p$, and whether the fermions are relativistic or non-relativistic. Saturation branches correspond to $m_{\rm eff}\to0$ and lead to radiation-like scaling of the coupled sector, $\rho_\phi\sim\rho_\psi\sim a^{-4}$. Asymptotic branches instead drive the scalar toward $\phi=0$, where the fermions recover their bare mass and redshift as matter.
}
\label{tab:summary_main_results}
\end{table*}

\appendix

\section{Degenerate Fermi gas at zero temperature}
\label{app:fermi_gas}

In this appendix, we derive the fluid description of the fermionic sector used in the main text and collect the exact expressions for the degenerate Fermi gas at zero temperature. We also show how the scalar source term can be written in terms of $\rho_\psi$ and $p_\psi$.

\subsection{Exact energy density and pressure}
\label{app:fermi_exact}

We now compute the exact thermodynamic quantities for a degenerate Fermi gas at zero temperature, where all momentum states are filled up to the Fermi momentum,
\begin{equation}
    p_F=(3\pi^2 n_\psi)^{1/3}.
\end{equation}
It is convenient to define
\begin{equation}
    N_\psi\equiv 3\pi^2 n_\psi,
\end{equation}
so that
\begin{equation}
    p_F=N_\psi^{1/3}.
\end{equation}

The exact energy density and pressure are
\begin{equation}
    \rho_\psi =  \frac{1}{\pi^2}\int_0^{p_F}dp\,p^2\sqrt{p^2+m_{\rm eff}^2},
    \label{eq:rho_psi_int_app}
\end{equation}
and
\begin{equation}
    p_\psi
    =
    \frac{1}{3\pi^2}
    \int_0^{p_F} dp\,\frac{p^4}{\sqrt{p^2+m_{\rm eff}^2}}.
    \label{eq:p_psi_int_app}
\end{equation}

Introducing the dimensionless ratio
\begin{equation}
    x\equiv \frac{m_{\rm eff}}{N_\psi^{1/3}}
    =\frac{m_{\rm eff}}{p_F},
\end{equation}
and changing integration variable to
\begin{equation}
    u\equiv \frac{p}{p_F},
\end{equation}
one obtains
\begin{equation}
    \rho_\psi=
    \frac{N_\psi^{4/3}}{\pi^2}
    \int_0^1 du\,u^2\sqrt{u^2+x^2},
\end{equation}
and
\begin{equation}
    p_\psi=
    \frac{N_\psi^{4/3}}{3\pi^2}
    \int_0^1 du\,\frac{u^4}{\sqrt{u^2+x^2}}.
\end{equation}
Performing the integrals gives
\begin{equation}
    \rho_\psi=
    \frac{N_\psi^{4/3}}{8\pi^2}\,F(x),
    \qquad
    p_\psi=
    \frac{N_\psi^{4/3}}{24\pi^2}\,P(x),
    \label{eq:rho_p_exact_app}
\end{equation}
where
\begin{equation}
    F(x)\equiv
    (2+x^2)\sqrt{1+x^2}
    -x^4\sinh^{-1}\!\left(\frac{1}{x}\right),
\end{equation}
and
\begin{equation}
    P(x)\equiv
    (2-3x^2)\sqrt{1+x^2}
    +3x^4\sinh^{-1}\!\left(\frac{1}{x}\right).
\end{equation}
The quantity entering the scalar equation is the trace combination
\begin{equation}
    \rho_\psi-3p_\psi.
\end{equation}
Using Eq.~\eqref{eq:rho_p_exact_app}, one finds
\begin{equation}
    \rho_\psi-3p_\psi
    =
    \frac{N_\psi^{4/3}}{8\pi^2}\bigl[F(x)-P(x)\bigr].
\end{equation}
Since
\begin{equation}
    F(x)-P(x)
    =
    4\left[
    x^2\sqrt{1+x^2}
    -x^4\sinh^{-1}\!\left(\frac{1}{x}\right)
    \right],
\end{equation}
this becomes
\begin{equation}
    \rho_\psi-3p_\psi
    =
    \frac{N_\psi^{4/3}}{2\pi^2}
    \left[
    x^2\sqrt{1+x^2}
    -x^4\sinh^{-1}\!\left(\frac{1}{x}\right)
    \right].
    \label{eq:rho_minus_3p_exact_app}
\end{equation}

Now, differentiating Eq.~\eqref{eq:rho_psi_int_app} with respect to $m_{\rm eff}$ gives
\begin{equation}
    \frac{\partial\rho_\psi}{\partial m_{\rm eff}}
    =
    \frac{m_{\rm eff}}{\pi^2}
    \int_0^{p_F} dp\,\frac{p^2}{\sqrt{p^2+m_{\rm eff}^2}}.
    \label{eq:drho_dm_app}
\end{equation}
On the other hand, using the identity
\begin{equation}
    p^4=p^2(p^2+m_{\rm eff}^2)-m_{\rm eff}^2p^2,
\end{equation}
the pressure integral may be rewritten as
\begin{align}
    p_\psi
    &=
    \frac{1}{3\pi^2}
    \int_0^{p_F} dp\,
    \frac{p^2(p^2+m_{\rm eff}^2)-m_{\rm eff}^2p^2}
    {\sqrt{p^2+m_{\rm eff}^2}}
    \notag\\
    &=
    \frac{1}{3\pi^2}
    \int_0^{p_F} dp\,p^2\sqrt{p^2+m_{\rm eff}^2}
    -
    \frac{m_{\rm eff}^2}{3\pi^2}
    \int_0^{p_F} dp\,\frac{p^2}{\sqrt{p^2+m_{\rm eff}^2}}
    \notag\\
    &=
    \frac{\rho_\psi}{3}
    -
    \frac{m_{\rm eff}^2}{3\pi^2}
    \int_0^{p_F} dp\,\frac{p^2}{\sqrt{p^2+m_{\rm eff}^2}}.
\end{align}
Therefore
\begin{equation}
    \rho_\psi-3p_\psi
    =
    \frac{m_{\rm eff}^2}{\pi^2}
    \int_0^{p_F} dp\,\frac{p^2}{\sqrt{p^2+m_{\rm eff}^2}}.
    \label{eq:trace_integral_app}
\end{equation}
Comparing Eqs.~\eqref{eq:drho_dm_app} and \eqref{eq:trace_integral_app}, one obtains
\begin{equation}
    \frac{\partial\rho_\psi}{\partial m_{\rm eff}}
    =
    \frac{\rho_\psi-3p_\psi}{m_{\rm eff}},
\end{equation}
as claimed.

For later use, it is convenient to define the exact dimensionless function
\begin{equation}
    \mathcal T(x)\equiv
    x\sqrt{1+x^2}
    -x^3\sinh^{-1}\!\left(\frac{1}{x}\right).
    \label{eq:Tdef_app}
\end{equation}
Using $m_{\rm eff}=xN_\psi^{1/3}$, Eq.~\eqref{eq:rho_minus_3p_exact_app} becomes
\begin{equation}
    \frac{\rho_\psi-3p_\psi}{m_{\rm eff}}
    =
    \frac{N_\psi}{2\pi^2}\,\mathcal T(x).
    \label{eq:trace_over_m_exact_app}
\end{equation}
This is the exact form entering the source term in the scalar equation of motion in the main text.

\subsection{Relativistic and non-relativistic limits}

For $x\ll1$, one expands
\begin{equation}
    \sqrt{1+x^2}=1+\frac{x^2}{2}+\mathcal O(x^4),
\end{equation}
and
\begin{equation}
    \sinh^{-1}\!\left(\frac{1}{x}\right)
    =
    \ln\!\left(\frac{2}{x}\right)+\mathcal O(x^2).
\end{equation}
This gives
\begin{align}
    F(x) &= 2 + 2x^2 + \mathcal{O}\!\left(x^4\ln x\right)\,,\notag \\
    P(x) &= 2 - 2x^2 + \mathcal{O}\!\left(x^4\ln x\right).
\end{align}
Hence
\begin{equation}
    \rho_\psi
    =
    \frac{N_\psi^{4/3}}{4\pi^2}
    \left[
    1+x^2+\mathcal O(x^4\ln x)
    \right],
\end{equation}
\begin{equation}
    p_\psi
    =
    \frac{N_\psi^{4/3}}{12\pi^2}
    \left[
    1-x^2+\mathcal O(x^4\ln x)
    \right],
\end{equation}
and
\begin{equation}
    \rho_\psi-3p_\psi
    =
    \frac{N_\psi^{4/3}}{2\pi^2}
    \left[
    x^2+\mathcal O(x^4\ln x)
    \right].
    \label{eq:rho_minus_3p_rel_app}
\end{equation}
Equivalently,
\begin{equation}
    \frac{\rho_\psi-3p_\psi}{m_{\rm eff}}
    \simeq
    \frac{N_\psi^{2/3}m_{\rm eff}}{2\pi^2},
    \qquad x\ll1,
    \label{eq:trace_over_m_rel_app}
\end{equation}
or
\begin{equation}
    \mathcal T(x)=x+\mathcal O(x^3\ln x).
    \label{eq:T_rel_app}
\end{equation}

For $x\gg1$, one uses
\begin{equation}
    \sqrt{1+x^2}
    =
    x+\frac{1}{2x}+\mathcal O(x^{-3}),
\end{equation}
and
\begin{equation}
    \sinh^{-1}\!\left(\frac{1}{x}\right)
    =
    \frac{1}{x}-\frac{1}{6x^3}+\mathcal O(x^{-5}).
\end{equation}
This yields
\begin{equation}
    F(x)=\frac{8}{3}x+\frac{4}{5x}+\mathcal O(x^{-3}),
    \qquad
    P(x)=\frac{8}{5x}+\mathcal O(x^{-3}).
\end{equation}
Hence
\begin{equation}
    \rho_\psi
    =
    \frac{N_\psi^{4/3}}{3\pi^2}
    \left[
    x+\frac{3}{10x}+\mathcal O(x^{-3})
    \right],
\end{equation}
\begin{equation}
    p_\psi
    =
    \frac{N_\psi^{4/3}}{15\pi^2}
    \left[
    \frac{1}{x}+\mathcal O(x^{-3})
    \right],
\end{equation}
and
\begin{equation}
    \rho_\psi-3p_\psi
    =
    \frac{N_\psi^{4/3}}{3\pi^2}
    \left[
    x-\frac{3}{10x}+\mathcal O(x^{-3})
    \right].
    \label{eq:rho_minus_3p_nr_app}
\end{equation}
Therefore
\begin{equation}
    \frac{\rho_\psi-3p_\psi}{m_{\rm eff}}
    \simeq
    \frac{N_\psi}{3\pi^2},
    \qquad x\gg1,
    \label{eq:trace_over_m_nr_app}
\end{equation}
or equivalently
\begin{equation}
    \mathcal T(x)=\frac{2}{3}+\mathcal O(x^{-2}).
    \label{eq:T_nr_app}
\end{equation}

\section{Scale-factor form of the scalar equation}
\label{app:scale_factor}

We now rewrite the homogeneous scalar equation in terms of derivatives with respect to the scale factor $a$.

For any homogeneous scalar $\varphi(a)$, one has
\begin{equation}
    \dot \varphi(t)=\frac{\dd\varphi}{\dd a}\dot a=aH \varphi'(a)\,,
\end{equation}
where a prime denotes $d/da$. Therefore
\begin{equation}
    \dot\varphi=aH\varphi'.
\end{equation}
Differentiating once more,
\begin{align}
    \ddot\varphi
    &=
    \frac{\dd}{\dd t}(aH\varphi')
    \notag\\
    &=
    (\dot a H+a\dot H)\varphi'+aH\frac{\dd}{\dd t}\varphi'
    \notag\\
    &=
    a(H^2+\dot H)\varphi'+aH(aH\varphi'')
    \notag\\
    &=
    a^2H^2\varphi''+a(H^2+\dot H)\varphi'.
\end{align}

For a background dominated by a perfect fluid with constant equation-of-state parameter $w$,
\begin{equation}
    H^2=H_i^2a^{-3(1+w)}.
\end{equation}
Differentiating with respect to time,
\begin{equation}
    2H\dot H
    =
    -3(1+w)H_i^2a^{-3(1+w)-1}\dot a
    =
    -3(1+w)H^3,
\end{equation}
which implies
\begin{equation}
    \dot H=-\frac32(1+w)H^2.
\end{equation}
Substituting this into the previous expression yields
\begin{equation}
    \ddot\varphi
    =
    a^2H^2\varphi''
    -\frac{1+3w}{2}aH^2\varphi'.
    \label{eq:varphi_ddot_app}
\end{equation}

The homogeneous scalar equation in cosmic time is
\begin{equation}
    \ddot\varphi+3H\dot\varphi+V'_{\rm bare}(\varphi)+\mathcal S_\varphi=0,
\end{equation}
with
\begin{equation}
    \mathcal S_\varphi\equiv\frac{\partial\rho_\psi}{\partial\varphi}.
\end{equation}
Using $\dot\varphi=aH\varphi'$ and Eq.~\eqref{eq:varphi_ddot_app}, one finds
\begin{equation}
    a^2H^2\varphi''
    +\frac{5-3w}{2}aH^2\varphi'
    +\frac{\dd V_{\rm bare}(\varphi)}{\dd\varphi}+\mathcal S_\varphi=0.
\end{equation}
Dividing by $a^2H^2$ gives
\begin{equation}
    \varphi''+\frac{5-3w}{2a}\varphi'
    +\frac{1}{a^2H^2}\frac{\dd V_{\rm bare}(\varphi)}{\dd\varphi}
    +\frac{\mathcal S_\varphi}{a^2H^2}=0.
    \label{eq:scale_factor_eom_generic_app}
\end{equation}
This is the starting point for the dimensionless equations written in the main text.

\section{Parabolic waves in the non-relativistic regime}
\label{app:Parabolic_waves}
We briefly explain the origin of the parabolic-wave approximation used for the non-relativistic oscillations around the saturation point \cite{Zygmund2002}. As discussed in the main text, the non-relativistic equation of motion is, at leading order, cusp-like,
\begin{equation}
    u'' + \frac{5-3w}{2a}u' + Dp a^{3w-2}{\rm sgn}(u) = 0\,.
    \label{eq:parabolic_eom}
\end{equation}
The key point is that the force is approximately constant on either side of the cusp. Hence, the motion is not harmonic. Instead, for a fixed value of the slowly varying coefficient $Dp a^{3w-2}$, the solution is made of parabolic arcs. Let $U(a)$ denote the local oscillation amplitude
of $u$. The instantaneous "energy" of the cusp oscillator is
\begin{equation}
    E(a) = \frac{1}{2}u'^2 + F(a)|u|\,,
    \qquad
    F(a)=Dp a^{3w-2}.
\end{equation}
At a turning point, $E=F U$. Averaging Eq.~\eqref{eq:parabolic_eom}
over one oscillation gives
\begin{equation}
    \frac{\dd E}{\dd a} =
    -\frac{5-3w}{2a}\langle u'^2\rangle +
    F'(a)\langle |u| \rangle .
\end{equation}
For a linear cusp potential, the virial relation gives
\begin{equation}
   \frac{1}{2} \langle u'^2\rangle
    = \frac{E}{3},
    \qquad
   F(a)\langle |u| \rangle = \frac{2E}{3}\,.
\end{equation}
Therefore,
\begin{equation}
    \frac{d\ln E}{d\ln a}
    =
    3(w-1),
\end{equation}
and since $F(a)\propto a^{3w-2}$, the amplitude satisfies
\begin{equation}
    U(a)=\frac{E(a)}{F(a)}
    =
    \frac{A_p}{a}.
    \label{eq:parabolic_envelope}
\end{equation}
The constant $A_p$ is fixed by matching to the full solution and
depends on the initial conditions and on the transient evolution before the field settles into the saturation regime.

We now determine the leading phase. For a cusp oscillator with constant
force $F$ and amplitude $U$, the time in scale-factor variable from a
turning point to the first zero crossing is
\begin{equation}
    \Delta a_{\rm quarter}
    =
    \sqrt{\frac{2U}{F}}.
\end{equation}
Hence, the full period is
\begin{equation}
    T_a
    =
    4\sqrt{\frac{2U}{F}},
\end{equation}
and the instantaneous angular frequency is
\begin{equation}
    \frac{\dd\theta}{\dd a}
    =
    \frac{2\pi}{T_a}
    =
    \frac{\pi}{2\sqrt{2}}
    \sqrt{\frac{F(a)}{U(a)}}.
\end{equation}
Using $F(a)=Dp a^{3w-2}$ and $U(a)=A_p/a$, this gives
\begin{equation}
    \frac{\dd\theta}{\dd a}
    =
    \frac{\pi}{2\sqrt{2}}
    \sqrt{\frac{Dp}{A_p}}
    a^{(3w-1)/2}.
\end{equation}
Therefore, for $w\neq -1/3$,
\begin{equation}
    \theta(a) = \frac{\pi}{\sqrt{2}(3w+1)}
    \sqrt{\frac{Dp}{A_p}}
    a^{(3w+1)/2}
    +
    \Theta_p(a),
    \label{eq:parabolic_phase}
\end{equation}
where $\Theta_p(a)$ contains the constant phase and possible residual subleading phase corrections. Then, the leading asymptotic solution can be written as
\begin{equation}
    \phi(a) = -1 + \frac{A_p}{a}
    P\left[\theta(a)\right],
    \label{eq:parabolic_wave_solution}
\end{equation}
where $P$ is a unit-amplitude periodic parabolic wave with period
$2\pi$.

The explicit form of $P$ follows directly from the constant-force motion. We choose the phase convention such that $\theta=0$ corresponds to the positive turning point, $P=1$, and define
\begin{equation}
    z=\frac{\theta}{2\pi}
    \quad {\rm mod}\quad 1\,.
\end{equation}
During the first quarter period, $0\le z<1/4$, the displacement is positive and the acceleration is constant and negative. Therefore $P$ is a parabola satisfying $P(0)=1$, $dP/dz|_{z=0}=0$, and $P(1/4)=0$. During the next half period, $1/4\le z<3/4$, the displacement is negative and the acceleration changes sign. The corresponding branch is then fixed by continuity of $P$ and $dP/dz$ at $z=1/4$. Finally, the last branch is fixed by continuity at $z=3/4$ and by the periodicity condition $P(1)=P(0)=1$. This gives
\begin{equation}
P(\theta)=
\begin{cases}
1-16z^2,
&0\leq z<\frac{1}{4},\\
-8\left(z-\frac{1}{4}\right)
+16\left(z-\frac{1}{4}\right)^2,
&\frac{1}{4}\leq z<\frac{3}{4},\\
8\left(z-\frac{3}{4}\right)
-16\left(z-\frac{3}{4}\right)^2,
&\frac{3}{4}\leq z<1.
\end{cases}
\label{eq:parabolic_wave_piecewise}
\end{equation}

The Fourier expansion of the same waveform is
\begin{equation}
    P(\theta)
    =
    \frac{32}{\pi^3}
    \sum_{k=0}^{\infty}
    \frac{(-1)^k}{(2k+1)^3}
    \cos\left[(2k+1)\theta\right].
    \label{eq:parabolic_wave_fourier}
\end{equation}
Keeping only the first harmonic gives
\begin{equation}
    P(\theta)
    \simeq
    \frac{32}{\pi^3}\cos\theta.
\end{equation}
Therefore, the first-harmonic approximation to
Eq.~\eqref{eq:parabolic_wave_solution} is
\begin{equation}
    \phi(a)
    \simeq
    -1 + \frac{32A_p}{\pi^3 a} \cos\left(
        \sqrt{
        \frac{\pi^2 Dp}{2(3w+1)^2A_p}
        }
        a^{(3w+1)/2}
        +
        \Theta_p(a)
    \right).
    \label{eq:parabolic_first_harmonic}
\end{equation}
This is the expression used in the main text. The cosine approximation should therefore be understood as the first harmonic of the underlying piecewise-parabolic waveform. The non-harmonicity is encoded in the higher Fourier modes of $P$, while the leading envelope is $|\phi+1|\sim a^{-1}$.

\section{Dilatonic coupling} \label{app:dilaton_scalings}

In this appendix, we supplement the discussion in the main text with a detailed study on a bare fermion mass and the dilatonic coupling 
\begin{equation}\label{eq:fvarphidilaton}
    f(\varphi)=\pm M_{\rm pl} e^{c\varphi/M_{\rm pl}}\,.
\end{equation}
Similarly to the procedure we followed for the monomial case (see Sec.~\ref{sec:representative_couplings}), we define the dimensionless quantities,
\begin{equation}
    \phi\equiv \frac{c\varphi}{M_{\rm pl}},
    \qquad
    B\equiv \frac{y M_{\rm pl}}{m_\psi}>0,
\end{equation}
so that the effective fermion mass reads
\begin{equation}
    m_{\rm eff}=m_\psi\left|1\pm B e^\phi\right|.
\end{equation}
Similarly, the ratio $x$ is now given by 
\begin{equation}
    x=x_ia
    \frac{|1\pm B e^\phi|}{|1\pm B e^{\phi_i}|}\,.
\end{equation}
We then obtain the scalar equations of motion \eqref{eq:phi_exp_rel_main_new} in the \textit{relativistic regime} $x\ll1$:
\begin{align}
    &\phi''+\frac{5-3w}{2a}\phi'
    +\widetilde{\Lambda}_q a^{3w+1}\phi^{q-1}\notag\\
    &\qquad\qquad\pm K_{\rm exp}a^{3w-1}
    e^\phi(1\pm B e^\phi)=0\,.
    \label{eq:phi_exp_rel_main_new}
\end{align}
while it reduces to
\begin{align}
    &\phi''+\frac{5-3w}{2a}\phi'
    +\widetilde{\Lambda}_q a^{3w+1}\phi^{q-1}\notag\\
    &\qquad\qquad\pm D_{\rm exp}a ^{3w-2}
    e^\phi\,\sgn(1\pm B e^\phi)=0,
    \label{eq:phi_exp_nr_main_new}
\end{align}
in the \textit{non-relativistic regime}, that is $x\gg1$. In Eqs.~\eqref{eq:phi_exp_rel_main_new} and \eqref{eq:phi_exp_nr_main_new}, we respectively introduced
\begin{equation}
    K_{\rm exp}\equiv
    \frac{6\,c^2 B x_i^2}
    {|1\pm B e^{\phi_i}|^2}\,\Omega_{\psi,i}
     \,\,,\,\,
    D_{\rm exp}\equiv
    \frac{3c^2 B }
    {|1\pm B e^{\phi_i}|}\,\Omega_{\psi,i}\,,
\end{equation}
and 
\begin{align}
\widetilde{\Lambda}_q\equiv
    \frac{\lambda_q M_{\rm pl}^{q-2}}{H_i^2 c^{\,q-2}}\,,
\end{align}
coming from the bare scalar potential, for compactness.

\subsection{Dilatonic coupling with $V_\text{bare}=0$}
\label{sec:dilatonic_results}

From now on, let us focus on the $-$ sign in Eq.~\eqref{eq:fvarphidilaton}, since in this case there exists a saturation point with $m_{\rm eff}=0$. 
Ignoring the bare potential, in the relativistic regime, the effective potential we obtain is
\begin{equation}
    V_{\rm eff}^{\rm rel}(\phi\,;\,a)
    =
    K_{\rm exp}
    \left(
    -e^\phi+\frac{B}{2}e^{2\phi}
    \right)a^{3w-1}.
\end{equation}
with its minimum at $\phi_s=-\ln B$. Hence, again the relativistic dynamics drives the field toward the point where the effective fermion mass vanishes.

Expanding near the minimum with $u\equiv \phi-\phi_s$, and using $e^{\phi_s}=1/B$,
the equation becomes
\begin{equation}
    u''+\frac{5-3w}{2a}u'
    +\frac{K_{\rm exp}}{B}a^{3w-1}u=0.
\end{equation}
Thus, for $w\neq -1/3$, the late-time solution then is given by
\begin{equation}
    \phi(a)\simeq
    \phi_s+\frac{A}{a}
    \cos\left(
    \frac{2\sqrt{K_{\rm exp}/B}}{3w+1}
    a^{\frac{3w+1}{2}}+\theta
    \right).
\end{equation}
This is qualitatively identical to the monomial case. Hence, the energy density of the scalar at the oscillatory regime behaves again as a radiation fluid and recall that the relativistic fermions behave in the same manner, so the two sectors track each other $\rho_\phi\sim\rho_\psi\sim a^{-4}$.

In the non-relativistic regime, the effective potential becomes
\begin{equation}
    V_{\rm eff}^{\text{non-rel}}(\phi\,;\,a)
    =
    \frac{D_{\rm exp}}{B}a^{3w-2}|1-B e^\phi|.
\end{equation}
As in the monomial odd-$p$ case, the solutions around $\phi_s$ are non-harmonic oscillations with a displacement envelope scaling as $a^{-1}$. Consequently, $|1-Be^\phi|\sim a^{-1}$ and the fermion energy density scales as $\rho_\psi\sim a^{-4}$.
Interestingly, for $w=1/3$ and especially for $w=0$--where the effective potential decays-- we have found numerically scenarios in which, starting with $\phi>\phi_s$, the field descends rapidly towards negative values and gets stuck there, where the exponential term dies off.

\subsection{Asymptotic scalings for the dilatonic coupling}

We collect here the expected late-time behaviour for the purely
dilatonic coupling with a vanishing bare scalar potential.
\begin{equation}
    f(\varphi)=M_{\rm pl} e^{c\varphi/M_{\rm pl}}\,,
\end{equation}
with $y>0$.
For the massless fermion case considered here, the effective fermion mass is therefore $m_{\rm eff}(\varphi)=|f(\varphi)|$.
Defining the dimensionless field as earlier,
\begin{equation}
    \phi=\frac{c\varphi}{M_{\rm pl}},
    \qquad
    m_{\rm eff}=y M_{\rm pl} e^\phi.
\end{equation}
We take $c>0$, so that the weak-coupling direction corresponds to $\phi\to-\infty$. For $c<0$, the direction of motion is reversed.

In the equations below, the positive constants $K$ and $D$ absorb the appropriate powers of $yM_{\rm pl}$, together with the normalisation factors arising from the field redefinition. With this convention, the scale-factor dependence of the equations is unchanged.

In the relativistic regime, $x\ll1$, the leading mass-dependent
correction to the fermion energy density scales as $m_{\rm eff}^2$.
The scalar equation, therefore takes the form
\begin{equation}
    \phi'' + \frac{5-3w}{2a}\phi' + 2K a^{3w-1}e^{2\phi} =0\,,
    \qquad
    x\ll1\,.
\end{equation}
For $-1/3<w<1$, this admits the late-time scaling solution
\begin{equation}
    \phi_{\rm rel}(a) \simeq c_{{\rm rel},0} - \frac{3w+1}{2}\ln a,
\end{equation}
where the constant is fixed by
\begin{equation}
   c_{{\rm rel},0} = \frac{1}{2}\ln\frac{3(1-w)(3w+1)}{8K}\,.
\end{equation}
Thus unlike the monomial coupling apart from $w=1/3$ the universal envelope $a^{-1}$ for the effective mass does not hold, instead,
\begin{equation}
    m_{\rm eff}\propto a^{-(3w+1)/2}\,.
\end{equation}
For kination, $w=1$, the friction term no longer fixes a universal
amplitude. Writing the asymptotic solution as
\begin{equation}
    \phi_{\rm rel}(a) \simeq - \left(2+\sqrt{2E_{\rm rel}}\right)\ln a\,,
\end{equation}
the constant $E_{\rm rel}$ depends on $K$ and on the initial scalar
data. Equivalently, the $w=1$ equation admits the first integral which is conserved and can be determined by the initial conditions:
\begin{equation}
    E_{\rm rel} = \frac{1}{2}\left(a\phi_i' +2\right)^2 + K a_i^4 e^{2\phi_i}\,,
\end{equation}
which can be fixed by the initial conditions. The main result of this analysis is that for purely dilatonic coupling for $w<1$ there is a true scaling solution which is an attractor and independent of the initial data, unlike the monomial case where the amplitude is dependent on the initial data. For $w=1$ on the other hand a memory of the initial data remains through $E_{\rm rel}$.

A similar pattern is found in the non-relativistic regime, $x\gg1$, where the fermion energy density is
linear in the mass, and the scalar obeys
\begin{equation}
    \phi'' + \frac{5-3w}{2a}\phi' + D a^{3w-2}e^\phi = 0\,,
    \qquad x\gg1.
\end{equation}
For $0<w<1$, the late-time attractor is
\begin{equation}
    \phi_{\rm nr}(a) \simeq c_{{\rm nr},0} - 3w\ln a\,,
\end{equation}
with
\begin{equation}
    c_{{\rm nr},0} = \ln\frac{9w(1-w)}{2D}\,,
\end{equation}
and therefore $m_{\rm eff}\propto a^{-3w}$, so that the non-relativistic fermion energy density tracks the background as $\rho_\psi\propto a^{-3(1+w)}$.

The matter-dominated case $w=0$ is marginal and instead gives
\begin{equation}
    \phi_{\rm nr}(a)
    \simeq
    -\ln\left(\ln a\right)
    +
    \ln\left(\frac{3}{2D}\right).
\end{equation}
For kination, $w=1$, the asymptotic behaviour is again not universal:
\begin{equation}
    \phi_{\rm nr}(a)\simeq - \left(3+\sqrt{2E_{\rm nr}}\right)\ln a,
\end{equation}
where
\begin{equation}
    E_{\rm nr} = \frac{1}{2}\left(a\phi_i' +3\right)^2 + D a_i^3 e^\phi_i\,.
\end{equation}

\bibliographystyle{apsrev4-1}
\bibliography{mybib.bib}

@article{Bowman:2018yin,
    author = "Bowman, Judd D. and Rogers, Alan E. E. and Monsalve, Raul A. and Mozdzen, Thomas J. and Mahesh, Nivedita",
    title = "{An absorption profile centred at 78 megahertz in the sky-averaged spectrum}",
    eprint = "1810.05912",
    archivePrefix = "arXiv",
    primaryClass = "astro-ph.CO",
    doi = "10.1038/nature25792",
    journal = "Nature",
    volume = "555",
    number = "7694",
    pages = "67--70",
    year = "2018"
}

@article{Lovell_2022,
   title={Extreme value statistics of the halo and stellar mass distributions at high redshift: are <i>JWST</i> results in tension with ΛCDM?},
   volume={518},
   ISSN={1365-2966},
   url={http://dx.doi.org/10.1093/mnras/stac3224},
   DOI={10.1093/mnras/stac3224},
   number={2},
   journal={Monthly Notices of the Royal Astronomical Society},
   publisher={Oxford University Press (OUP)},
   author={Lovell, Christopher C and Harrison, Ian and Harikane, Yuichi and Tacchella, Sandro and Wilkins, Stephen M},
   year={2022},
   month=Nov, pages={2511–2520} }

@article{Boylan-Kolchin:2022kae,
    author = "Boylan-Kolchin, Michael",
    title = "{Stress testing {\ensuremath{\Lambda}}CDM with high-redshift galaxy candidates}",
    eprint = "2208.01611",
    archivePrefix = "arXiv",
    primaryClass = "astro-ph.CO",
    doi = "10.1038/s41550-023-01937-7",
    journal = "Nature Astron.",
    volume = "7",
    number = "6",
    pages = "731--735",
    year = "2023"
}

@inbook{Haiman_2012,
   title={The Formation of the First Massive Black Holes},
   ISBN={9783642323621},
   ISSN={0067-0057},
   url={http://dx.doi.org/10.1007/978-3-642-32362-1_6},
   DOI={10.1007/978-3-642-32362-1_6},
   booktitle={The First Galaxies},
   publisher={Springer Berlin Heidelberg},
   author={Haiman, Zoltán},
   year={2012},
   month=Sept, pages={293–341} }

@misc{shankaranarayanan2026primordialblackholesreview,
      title={Primordial Black Holes: A Review of Formation and Evolution}, 
      author={S. Shankaranarayanan and Soumya Bhattacharya and Archit Vidyarthi},
      year={2026},
      eprint={2606.23846},
      archivePrefix={arXiv},
      primaryClass={gr-qc},
      url={https://arxiv.org/abs/2606.23846}, 
}

@article{Azhar:2018lzd,
    author = "Azhar, Feraz and Loeb, Abraham",
    title = "{Gauging Fine-Tuning}",
    eprint = "1809.06220",
    archivePrefix = "arXiv",
    primaryClass = "astro-ph.CO",
    doi = "10.1103/PhysRevD.98.103018",
    journal = "Phys. Rev. D",
    volume = "98",
    number = "10",
    pages = "103018",
    year = "2018"
}

@article{Hertzberg:2017dkh,
    author = "Hertzberg, Mark P. and Yamada, Masaki",
    title = "{Primordial Black Holes from Polynomial Potentials in Single Field Inflation}",
    eprint = "1712.09750",
    archivePrefix = "arXiv",
    primaryClass = "astro-ph.CO",
    doi = "10.1103/PhysRevD.97.083509",
    journal = "Phys. Rev. D",
    volume = "97",
    number = "8",
    pages = "083509",
    year = "2018"
}

@article{Nakama:2018utx,
    author = "Nakama, Tomohiro and Wang, Yi",
    title = "{Do we need fine-tuning to create primordial black holes?}",
    eprint = "1811.01126",
    archivePrefix = "arXiv",
    primaryClass = "astro-ph.CO",
    doi = "10.1103/PhysRevD.99.023504",
    journal = "Phys. Rev. D",
    volume = "99",
    number = "2",
    pages = "023504",
    year = "2019"
}

@article{Allahverdi:2020bys,
    author = "Allahverdi, Rouzbeh and others",
    title = "{The First Three Seconds: a Review of Possible Expansion Histories of the Early Universe}",
    eprint = "2006.16182",
    archivePrefix = "arXiv",
    primaryClass = "astro-ph.CO",
    reportNumber = "FERMILAB-PUB-20-242-A, KCL-PH-TH/2020-33, KEK-Cosmo-257,
  KEK-TH-2231, IPMU20-0070, PI/UAN-2020-674FT, RUP-20-22",
    doi = "10.21105/astro.2006.16182",
    journal = "Open J. Astrophys.",
    volume = "4",
    pages = "astro.2006.16182",
    year = "2021"
}

@article{Braglia:2022phb,
    author = "Braglia, Matteo and Linde, Andrei and Kallosh, Renata and Finelli, Fabio",
    title = "{Hybrid {\ensuremath{\alpha}}-attractors, primordial black holes and gravitational wave backgrounds}",
    eprint = "2211.14262",
    archivePrefix = "arXiv",
    primaryClass = "astro-ph.CO",
    doi = "10.1088/1475-7516/2023/04/033",
    journal = "JCAP",
    volume = "04",
    pages = "033",
    year = "2023"
}

@article{Carr:2019hud,
    author = "Carr, Bernard and Clesse, Sebastien and Garc{\'\i}a-Bellido, Juan",
    title = "{Primordial black holes from the QCD epoch: Linking dark matter, baryogenesis and anthropic selection}",
    eprint = "1904.02129",
    archivePrefix = "arXiv",
    primaryClass = "astro-ph.CO",
    doi = "10.1093/mnras/staa3726",
    journal = "Mon. Not. Roy. Astron. Soc.",
    volume = "501",
    number = "1",
    pages = "1426--1439",
    year = "2021"
}

@article{Qin:2023lgo,
    author = "Qin, Wenzer and Geller, Sarah R. and Balaji, Shyam and McDonough, Evan and Kaiser, David I.",
    title = "{Planck constraints and gravitational wave forecasts for primordial black hole dark matter seeded by multifield inflation}",
    eprint = "2303.02168",
    archivePrefix = "arXiv",
    primaryClass = "astro-ph.CO",
    reportNumber = "MIT-CTP/5525",
    doi = "10.1103/PhysRevD.108.043508",
    journal = "Phys. Rev. D",
    volume = "108",
    number = "4",
    pages = "043508",
    year = "2023"
}

@article{Animali:2022otk,
    author = "Animali, Chiara and Vennin, Vincent",
    title = "{Primordial black holes from stochastic tunnelling}",
    eprint = "2210.03812",
    archivePrefix = "arXiv",
    primaryClass = "astro-ph.CO",
    doi = "10.1088/1475-7516/2023/02/043",
    journal = "JCAP",
    volume = "02",
    pages = "043",
    year = "2023"
}

@article{Brax:2010gi,
  author = {Brax, Philippe and van de Bruck, Carsten and Davis, Anne-Christine and Shaw, Douglas J.},
  title = {The Dilaton and Modified Gravity},
  eprint = {1005.3735},
  archivePrefix = {arXiv},
  primaryClass = {astro-ph.CO},
  journal = {Phys. Rev. D},
  volume = {82},
  pages = {063519},
  year = {2010},
  doi = {10.1103/PhysRevD.82.063519}
}

@article{Brax_2022,
   title={Cointeracting dark matter and conformally coupled light scalars},
   volume={105},
   ISSN={2470-0029},
   url={http://dx.doi.org/10.1103/PhysRevD.105.103015},
   DOI={10.1103/physrevd.105.103015},
   number={10},
   journal={Physical Review D},
   publisher={American Physical Society (APS)},
   author={Brax, Philippe and van de Bruck, Carsten and Trojanowski, Sebastian},
   year={2022},
   month=May }

@article{Zeldovich:1967lct,
    author = "Zel'dovich, Ya. B. and Novikov, I. D.",
    title = "{The Hypothesis of Cores Retarded during Expansion and the Hot Cosmological Model}",
    journal = "Sov. Astron.",
    volume = "10",
    pages = "602",
    year = "1967"
}

@article{Hawking:1971ei,
    author = "Hawking, Stephen",
    title = "{Gravitationally collapsed objects of very low mass}",
    doi = "10.1093/mnras/152.1.75",
    journal = "Mon. Not. Roy. Astron. Soc.",
    volume = "152",
    pages = "75",
    year = "1971"
}

@article{Carr:1974nx,
    author = "Carr, Bernard J. and Hawking, S. W.",
    title = "{Black holes in the early Universe}",
    doi = "10.1093/mnras/168.2.399",
    journal = "Mon. Not. Roy. Astron. Soc.",
    volume = "168",
    pages = "399--415",
    year = "1974"
}

@article{Amendola:2017xhl,
    author = "Amendola, Luca and Rubio, Javier and Wetterich, Christof",
    title = "{Primordial black holes from fifth forces}",
    eprint = "1711.09915",
    archivePrefix = "arXiv",
    primaryClass = "astro-ph.CO",
    doi = "10.1103/PhysRevD.97.081302",
    journal = "Phys. Rev. D",
    volume = "97",
    number = "8",
    pages = "081302",
    year = "2018"
}

@article{Cole:2023wyx,
    author = "Cole, Philippa S. and Gow, Andrew D. and Byrnes, Christian T. and Patil, Subodh P.",
    title = "{Primordial black holes from single-field inflation: a fine-tuning audit}",
    eprint = "2304.01997",
    archivePrefix = "arXiv",
    primaryClass = "astro-ph.CO",
    doi = "10.1088/1475-7516/2023/08/031",
    journal = "JCAP",
    volume = "08",
    pages = "031",
    year = "2023"
}

@article{Ozsoy:2023ryl,
    author = {{\"O}zsoy, Ogan and Tasinato, Gianmassimo},
    title = "{Inflation and Primordial Black Holes}",
    eprint = "2301.03600",
    archivePrefix = "arXiv",
    primaryClass = "astro-ph.CO",
    doi = "10.3390/universe9050203",
    journal = "Universe",
    volume = "9",
    number = "5",
    pages = "203",
    year = "2023"
}

@article{Savastano:2019zpr,
    author = "Savastano, Stefano and Amendola, Luca and Rubio, Javier and Wetterich, Christof",
    title = "{Primordial dark matter halos from fifth forces}",
    eprint = "1906.05300",
    archivePrefix = "arXiv",
    primaryClass = "astro-ph.CO",
    reportNumber = "HIP-2019-18/TH",
    doi = "10.1103/PhysRevD.100.083518",
    journal = "Phys. Rev. D",
    volume = "100",
    number = "8",
    pages = "083518",
    year = "2019"
}

@article{Dom_nech_2021,
   title={Cosmology of strongly interacting fermions in the early universe},
   volume={2021},
   ISSN={1475-7516},
   url={http://dx.doi.org/10.1088/1475-7516/2021/06/030},
   DOI={10.1088/1475-7516/2021/06/030},
   number={06},
   journal={Journal of Cosmology and Astroparticle Physics},
   publisher={IOP Publishing},
   author={Domènech, Guillem and Sasaki, Misao},
   year={2021},
   month=June, pages={030} }

@article{Dom_nech_2023,
   title={Halo formation from Yukawa forces in the very early Universe},
   volume={108},
   ISSN={2470-0029},
   url={http://dx.doi.org/10.1103/PhysRevD.108.103543},
   DOI={10.1103/physrevd.108.103543},
   number={10},
   journal={Physical Review D},
   publisher={American Physical Society (APS)},
   author={Domènech, Guillem and Inman, Derek and Kusenko, Alexander and Sasaki, Misao},
   year={2023},
   month=Nov }

@article{Amendola:1999er,
    author = "Amendola, Luca",
    title = "{Coupled quintessence}",
    eprint = "astro-ph/9908023",
    archivePrefix = "arXiv",
    doi = "10.1103/PhysRevD.62.043511",
    journal = "Phys. Rev. D",
    volume = "62",
    pages = "043511",
    year = "2000"
}

@article{Wetterich_1988,
   title={Cosmology and the fate of dilatation symmetry},
   volume={302},
   ISSN={0550-3213},
   url={http://dx.doi.org/10.1016/0550-3213(88)90193-9},
   DOI={10.1016/0550-3213(88)90193-9},
   number={4},
   journal={Nuclear Physics B},
   publisher={Elsevier BV},
   author={Wetterich, C.},
   year={1988},
   pages={668–696} }

@article{Farrar_2004,
   title={Interacting Dark Matter and Dark Energy},
   volume={604},
   ISSN={1538-4357},
   url={http://dx.doi.org/10.1086/381728},
   DOI={10.1086/381728},
   number={1},
   journal={The Astrophysical Journal},
   publisher={American Astronomical Society},
   author={Farrar, Glennys R. and Peebles, P. J. E.},
   year={2004},
   month=Mar, pages={1–11} }

@misc{baumann2014inflationstringtheory,
      title={Inflation and String Theory}, 
      author={Daniel Baumann and Liam McAllister},
      year={2014},
      eprint={1404.2601},
      archivePrefix={arXiv},
      primaryClass={hep-th},
      url={https://arxiv.org/abs/1404.2601}, 
}

@misc{cheng2026nonlineardynamicsprimordialblack,
      title={Non-linear Dynamics and Primordial Black Hole Formation During Kination}, 
      author={Cheng Cheng and Panagiotis Giannadakis and Lucien Heurtier and Eugene A. Lim},
      year={2026},
      eprint={2507.19166},
      archivePrefix={arXiv},
      primaryClass={astro-ph.CO},
      url={https://arxiv.org/abs/2507.19166}, 
}

@article{gouttenoire2022kinationcosmologyscalarfields,
      title={Kination cosmology from scalar fields and gravitational-wave signatures}, 
      author={Yann Gouttenoire and Geraldine Servant and Peera Simakachorn},
      year={2022},
      eprint={2111.01150},
      archivePrefix={arXiv},
      primaryClass={hep-ph},
      url={https://arxiv.org/abs/2111.01150}, 
}

@book{Fujii:2003pa,
    author = "Fujii, Y. and Maeda, K.",
    title = "{The scalar-tensor theory of gravitation}",
    doi = "10.1017/CBO9780511535093",
    isbn = "978-0-521-03752-5, 978-0-521-81159-0, 978-0-511-02988-2",
    publisher = "Cambridge University Press",
    series = "Cambridge Monographs on Mathematical Physics",
    month = "7",
    year = "2007"
}

@article{Jordan1959,
  author  = {Jordan, Pascual},
  title   = {The Present State of Dirac's Cosmological Hypothesis},
  journal = {Zeitschrift f{\"u}r Physik},
  volume  = {157},
  pages   = {112--121},
  year    = {1959}
}

@article{BransDicke1961,
  author  = {Brans, Carl H. and Dicke, Robert H.},
  title   = {Mach's Principle and a Relativistic Theory of Gravitation},
  journal = {Physical Review},
  volume  = {124},
  pages   = {925--935},
  year    = {1961}
}

@article{Amendola_2018,
   title={Primordial black holes from fifth forces},
   volume={97},
   ISSN={2470-0029},
   url={http://dx.doi.org/10.1103/PhysRevD.97.081302},
   DOI={10.1103/physrevd.97.081302},
   number={8},
   journal={Physical Review D},
   publisher={American Physical Society (APS)},
   author={Amendola, Luca and Rubio, Javier and Wetterich, Christof},
   year={2018},
   month=Apr }

@misc{apers2024stringtheoryhalfuniverse,
      title={String Theory and the First Half of the Universe}, 
      author={Fien Apers and Joseph P. Conlon and Edmund J. Copeland and Martin Mosny and Filippo Revello},
      year={2024},
      eprint={2401.04064},
      archivePrefix={arXiv},
      primaryClass={hep-th},
      url={https://arxiv.org/abs/2401.04064}, 
}

@article{Sasaki_2018,
   title={Primordial black holes—perspectives in gravitational wave astronomy},
   volume={35},
   ISSN={1361-6382},
   url={http://dx.doi.org/10.1088/1361-6382/aaa7b4},
   DOI={10.1088/1361-6382/aaa7b4},
   number={6},
   journal={Classical and Quantum Gravity},
   publisher={IOP Publishing},
   author={Sasaki, Misao and Suyama, Teruaki and Tanaka, Takahiro and Yokoyama, Shuichiro},
   year={2018},
   month=Feb, pages={063001} }

@article{Carr_2020,
   title={Primordial Black Holes as Dark Matter: Recent Developments},
   volume={70},
   ISSN={1545-4134},
   url={http://dx.doi.org/10.1146/annurev-nucl-050520-125911},
   DOI={10.1146/annurev-nucl-050520-125911},
   number={1},
   journal={Annual Review of Nuclear and Particle Science},
   publisher={Annual Reviews},
   author={Carr, Bernard and Kühnel, Florian},
   year={2020},
   month=Oct, pages={355–394} }

@article{Mosny:2025cyd,
    author = "Mosny, Martin and Conlon, Joseph P. and Copeland, Edmund J.",
    title = "{Self-tracking solutions for asymptotic scalar fields}",
    eprint = "2507.04161",
    archivePrefix = "arXiv",
    primaryClass = "hep-th",
    doi = "10.1007/JHEP12(2025)135",
    journal = "JHEP",
    volume = "12",
    pages = "135",
    year = "2025"
}

@article{Hall:2009bx,
    author = "Hall, Lawrence J. and Jedamzik, Karsten and March-Russell, John and West, Stephen M.",
    title = "{Freeze-In Production of FIMP Dark Matter}",
    eprint = "0911.1120",
    archivePrefix = "arXiv",
    primaryClass = "hep-ph",
    doi = "10.1007/JHEP03(2010)080",
    journal = "JHEP",
    volume = "03",
    pages = "080",
    year = "2010"
}

@article{Berlin:2016gtr,
    author = "Berlin, Asher and Hooper, Dan and Krnjaic, Gordan",
    title = "{PeV-Scale Dark Matter as a Thermal Relic of a Decoupled Sector}",
    eprint = "1602.08490",
    archivePrefix = "arXiv",
    primaryClass = "hep-ph",
    doi = "10.1016/j.physletb.2016.05.037",
    journal = "Phys. Lett. B",
    volume = "760",
    pages = "106--111",
    year = "2016"
}

@article{Berlin:2016vnh,
    author = "Berlin, Asher and Hooper, Dan and Krnjaic, Gordan",
    title = "{Thermal Dark Matter From A Highly Decoupled Sector}",
    eprint = "1609.02555",
    archivePrefix = "arXiv",
    primaryClass = "hep-ph",
    doi = "10.1103/PhysRevD.94.095019",
    journal = "Phys. Rev. D",
    volume = "94",
    number = "9",
    pages = "095019",
    year = "2016"
}

@article{Heurtier:2019beu,
    author = "Heurtier, Lucien and Huang, Fei and Toma, Takashi",
    title = "{The Inflaton Portal to a Highly Decoupled EeV Dark Matter Particle}",
    eprint = "1905.05191",
    archivePrefix = "arXiv",
    primaryClass = "hep-ph",
    doi = "10.1103/PhysRevD.100.043507",
    journal = "Phys. Rev. D",
    volume = "100",
    number = "4",
    pages = "043507",
    year = "2019"
}

@article{Garcia:2020eof,
    author = "Garcia, Marcos A. G. and Kaneta, Kunio and Mambrini, Yann and Olive, Keith A.",
    title = "{Reheating and Post-inflationary Production of Dark Matter}",
    eprint = "2004.08404",
    archivePrefix = "arXiv",
    primaryClass = "hep-ph",
    doi = "10.1103/PhysRevD.101.123507",
    journal = "Phys. Rev. D",
    volume = "101",
    number = "12",
    pages = "123507",
    year = "2020"
}

@article{Gubser:2004uh,
    author = "Gubser, Steven S. and Peebles, P. J. E.",
    title = "{Structure formation in a string inspired modification of the cold dark matter model}",
    eprint = "hep-th/0402225",
    archivePrefix = "arXiv",
    doi = "10.1103/PhysRevD.70.123510",
    journal = "Phys. Rev. D",
    volume = "70",
    pages = "123510",
    year = "2004"
}

@article{Nusser:2004qu,
    author = "Nusser, Adi and Gubser, Steven S. and Peebles, P. J. E.",
    title = "{Structure formation with a long-range scalar dark matter interaction}",
    eprint = "astro-ph/0412586",
    archivePrefix = "arXiv",
    doi = "10.1103/PhysRevD.71.083505",
    journal = "Phys. Rev. D",
    volume = "71",
    pages = "083505",
    year = "2005"
}

@article{Bean:2008ac,
    author = "Bean, Rachel and Flanagan, Eanna E. and Laszlo, Istvan and Trodden, Mark",
    title = "{Constraining interactions in cosmology's dark sector}",
    eprint = "0808.1105",
    archivePrefix = "arXiv",
    primaryClass = "astro-ph",
    doi = "10.1103/PhysRevD.78.123514",
    journal = "Phys. Rev. D",
    volume = "78",
    pages = "123514",
    year = "2008"
}

@article{Flores:2020drq,
    author = "Flores, Marcos M. and Kusenko, Alexander",
    title = "{Primordial Black Holes from Long-Range Scalar Forces and Scalar Radiative Cooling}",
    eprint = "2008.12456",
    archivePrefix = "arXiv",
    primaryClass = "astro-ph.CO",
    doi = "10.1103/PhysRevLett.126.041101",
    journal = "Phys. Rev. Lett.",
    volume = "126",
    number = "4",
    pages = "041101",
    year = "2021"
}

@article{Flores:2023zpf,
    author = "Flores, Marcos M. and Lu, Yifan and Kusenko, Alexander",
    title = "{Structure Formation after Reheating: Supermassive Primordial Black Holes and Fermi Ball Dark Matter}",
    eprint = "2308.09094",
    archivePrefix = "arXiv",
    primaryClass = "astro-ph.CO",
    doi = "10.1103/PhysRevD.108.123511",
    journal = "Phys. Rev. D",
    volume = "108",
    number = "12",
    pages = "123511",
    year = "2023"
}

@article{Kawana:2021tde,
    author = "Kawana, Kiyoharu and Xie, Ke-Pan",
    title = "{Primordial black holes from a cosmic phase transition: The collapse of Fermi-balls}",
    eprint = "2106.00111",
    archivePrefix = "arXiv",
    primaryClass = "astro-ph.CO",
    doi = "10.1016/j.physletb.2021.136791",
    journal = "Phys. Lett. B",
    volume = "824",
    pages = "136791",
    year = "2022"
}

@article{Lu:2024jzs,
    author = "Lu, Yifan and Picker, Zachary S. C. and Profumo, Stefano and Kusenko, Alexander",
    title = "{Black Holes from Fermi Ball Collapse}",
    eprint = "2411.17074",
    archivePrefix = "arXiv",
    primaryClass = "astro-ph.CO",
    doi = "10.1103/PhysRevD.111.043005",
    journal = "Phys. Rev. D",
    volume = "111",
    number = "4",
    pages = "043005",
    year = "2025"
}

@article{Graham:2025gtd,
    author = "Graham, Peter W. and Ramani, Harikrishnan and Simon, Olivier and Tanin, Erwin H.",
    title = "{Cosmological Limits on Strong Dark Forces}",
    eprint = "2511.09614",
    archivePrefix = "arXiv",
    primaryClass = "hep-ph",
    month = "11",
    year = "2025"
}

@article{Archidiacono:2022iuu,
    author = "Archidiacono, Maria and Castorina, Emanuele and Redigolo, Diego and Salvioni, Ennio",
    title = "{Unveiling dark fifth forces with linear cosmology}",
    eprint = "2204.08484",
    archivePrefix = "arXiv",
    primaryClass = "astro-ph.CO",
    reportNumber = "CERN-TH-2022-066",
    doi = "10.1088/1475-7516/2022/10/074",
    journal = "JCAP",
    volume = "10",
    pages = "074",
    year = "2022"
}

@article{Casas:2016duf,
    author = "Casas, Santiago and Pettorino, Valeria and Wetterich, Christof",
    title = "{Dynamics of neutrino lumps in growing neutrino quintessence}",
    eprint = "1608.02358",
    archivePrefix = "arXiv",
    primaryClass = "astro-ph.CO",
    doi = "10.1103/PhysRevD.94.103518",
    journal = "Phys. Rev. D",
    volume = "94",
    number = "10",
    pages = "103518",
    year = "2016"
}

@article{Wetterich:2007kr,
    author = "Wetterich, C.",
    title = "{Growing neutrinos and cosmological@misc{shankaranarayanan2026primordialblackholesreview,
      title={Primordial Black Holes: A Review of Formation and Evolution}, 
      author={S. Shankaranarayanan and Soumya Bhattacharya and Archit Vidyarthi},
      year={2026},
      eprint={2606.23846},
      archivePrefix={arXiv},
      primaryClass={gr-qc},
      url={https://arxiv.org/abs/2606.23846}, 
} selection}",
    eprint = "0706.4427",
    archivePrefix = "arXiv",
    primaryClass = "hep-ph",
    doi = "10.1016/j.physletb.2007.08.060",
    journal = "Phys. Lett. B",
    volume = "655",
    pages = "201--208",
    year = "2007"
}

@article{Amendola:2007yx,
    author = "Amendola, Luca and Baldi, Marco and Wetterich, Christof",
    title = "{Quintessence cosmologies with a growing matter component}",
    eprint = "0706.3064",
    archivePrefix = "arXiv",
    primaryClass = "astro-ph",
    doi = "10.1103/PhysRevD.78.023015",
    journal = "Phys. Rev. D",
    volume = "78",
    pages = "023015",
    year = "2008"
}

@article{Afshordi:2005ym,
    author = "Afshordi, Niayesh and Zaldarriaga, Matias and Kohri, Kazunori",
    title = "{On the stability of dark energy with mass-varying neutrinos}",
    eprint = "astro-ph/0506663",
    archivePrefix = "arXiv",
    doi = "10.1103/PhysRevD.72.065024",
    journal = "Phys. Rev. D",
    volume = "72",
    pages = "065024",
    year = "2005"
}

@article{Fardon:2003eh,
    author = "Fardon, Rob and Nelson, Ann E. and Weiner, Neal",
    title = "{Dark energy from mass varying neutrinos}",
    eprint = "astro-ph/0309800",
    archivePrefix = "arXiv",
    reportNumber = "UW-PT-03-22",
    doi = "10.1088/1475-7516/2004/10/005",
    journal = "JCAP",
    volume = "10",
    pages = "005",
    year = "2004"
}

@article{Wetterich:1994bg,
    author = "Wetterich, Christof",
    title = "{The Cosmon model for an asymptotically vanishing time dependent cosmological 'constant'}",
    eprint = "hep-th/9408025",
    archivePrefix = "arXiv",
    reportNumber = "HD-THEP-94-16",
    journal = "Astron. Astrophys.",
    volume = "301",
    pages = "321--328",
    year = "1995"
}

@article{Farrar:2003uw,
    author = "Farrar, Glennys R. and Peebles, P. James E.",
    title = "{Interacting dark matter and dark energy}",
    eprint = "astro-ph/0307316",
    archivePrefix = "arXiv",
    doi = "10.1086/381728",
    journal = "Astrophys. J.",
    volume = "604",
    pages = "1--11",
    year = "2004"
}

@article{Aurrekoetxea_2025,
   title={Cosmology using numerical relativity},
   volume={28},
   ISSN={1433-8351},
   url={http://dx.doi.org/10.1007/s41114-025-00058-z},
   DOI={10.1007/s41114-025-00058-z},
   number={1},
   journal={Living Reviews in Relativity},
   publisher={Springer Science and Business Media LLC},
   author={Aurrekoetxea, Josu C. and Clough, Katy and Lim, Eugene A.},
   year={2025},
   month=June }

@misc{dankovsky2025numericalanalysismeltingdomain,
      title={Numerical analysis of melting domain walls and their gravitational waves}, 
      author={I. Dankovsky and S. Ramazanov and E. Babichev and D. Gorbunov and A. Vikman},
      year={2025},
      eprint={2410.21971},
      archivePrefix={arXiv},
      primaryClass={hep-ph},
      url={https://arxiv.org/abs/2410.21971}, 
}

@article{Babichev_2023,
   title={NANOGrav spectral index 
<mml:math xmlns:mml=“http://www.w3.org/1998/Math/MathML” display=“inline”><mml:mi>γ</mml:mi><mml:mo>=</mml:mo><mml:mn>3</mml:mn></mml:math>
 from melting domain walls},
   volume={108},
   ISSN={2470-0029},
   url={http://dx.doi.org/10.1103/PhysRevD.108.123529},
   DOI={10.1103/physrevd.108.123529},
   number={12},
   journal={Physical Review D},
   publisher={American Physical Society (APS)},
   author={Babichev, E. and Gorbunov, D. and Ramazanov, S. and Samanta, R. and Vikman, A.},
   year={2023},
   month=Dec }

@book{Zygmund2002,
  author = {Zygmund, Antoni},
  title = {Trigonometric Series},
  edition = {Third},
  publisher = {Cambridge University Press},
  year = {2002}
}

@misc{colazo2024structureformationprimordialblack,
      title={Structure formation with primordial black holes to alleviate early star formation tension revealed by JWST}, 
      author={P. E. Colazo and F. Stasyszyn and N. Padilla},
      year={2024},
      eprint={2404.13110},
      archivePrefix={arXiv},
      primaryClass={astro-ph.CO},
      url={https://arxiv.org/abs/2404.13110}, 
}

@misc{hirano2024earlystructureformationprimordial,
      title={Early Structure Formation from Primordial Density Fluctuations with a Blue, Tilted Power Spectrum: High-Redshift Galaxies}, 
      author={Shingo Hirano and Naoki Yoshida},
      year={2024},
      eprint={2306.11993},
      archivePrefix={arXiv},
      primaryClass={astro-ph.GA},
      url={https://arxiv.org/abs/2306.11993}, 
}

@article{Inayoshi:2019fun,
  author = {Inayoshi, Kohei and Visbal, Eli and Haiman, Zoltan},
  title = {The Assembly of the First Massive Black Holes},
  eprint = {1911.05791},
  archivePrefix = {arXiv},
  primaryClass = {astro-ph.GA},
  journal = {Ann. Rev. Astron. Astrophys.},
  volume = {58},
  pages = {27--97},
  year = {2020}
}

@article{Dayal:2024zwq,
    author = "Dayal, Pratika",
    title = "{Exploring a primordial solution for early black holes detected with JWST}",
    eprint = "2407.07162",
    archivePrefix = "arXiv",
    primaryClass = "astro-ph.GA",
    doi = "10.1051/0004-6361/202451481",
    journal = "Astron. Astrophys.",
    volume = "690",
    pages = "A182",
    year = "2024"
}

@article{Del_Grosso_2023,
   title={Fermion soliton stars},
   volume={108},
   ISSN={2470-0029},
   url={http://dx.doi.org/10.1103/PhysRevD.108.044024},
   DOI={10.1103/physrevd.108.044024},
   number={4},
   journal={Physical Review D},
   publisher={American Physical Society (APS)},
   author={Del Grosso, Loris and Franciolini, Gabriele and Pani, Paolo and Urbano, Alfredo},
   year={2023},
   month=Aug }

@article{Lu_2025,
   title={Black holes from Fermi ball collapse},
   volume={111},
   ISSN={2470-0029},
   url={http://dx.doi.org/10.1103/PhysRevD.111.043005},
   DOI={10.1103/physrevd.111.043005},
   number={4},
   journal={Physical Review D},
   publisher={American Physical Society (APS)},
   author={Lu, Yifan and Picker, Zachary S. C. and Profumo, Stefano and Kusenko, Alexander},
   year={2025},
   month=Feb }

@article{Flores:2022uzt,
    author = "Flores, Marcos M. and Kusenko, Alexander and Sasaki, Misao",
    title = "{Gravitational Waves from Rapid Structure Formation on Microscopic Scales before Matter-Radiation Equality}",
    eprint = "2209.04970",
    archivePrefix = "arXiv",
    primaryClass = "astro-ph.CO",
    reportNumber = "IPMU22-0019, YITP-22-95",
    doi = "10.1103/PhysRevLett.131.011003",
    journal = "Phys. Rev. Lett.",
    volume = "131",
    number = "1",
    pages = "011003",
    year = "2023"
}

@article{Saikawa_2017,
   title={A Review of Gravitational Waves from Cosmic Domain Walls},
   volume={3},
   ISSN={2218-1997},
   url={http://dx.doi.org/10.3390/universe3020040},
   DOI={10.3390/universe3020040},
   number={2},
   journal={Universe},
   publisher={MDPI AG},
   author={Saikawa, Ken’ichi},
   year={2017},
   month=May, pages={40} }

\end{document}